\documentclass[]{pasj01}
\draft
\usepackage{lscape}

\Received{2020/01/24}
\Accepted{2020/02/07}
 
\begin{document} 

\title{ 
The First VERA Astrometry Catalog }

\author{VERA collaboration}
\author{Tomoya \textsc{Hirota}\altaffilmark{1,2}}
\email{tomoya.hirota@nao.ac.jp}
\author{Takumi \textsc{Nagayama}\altaffilmark{3}}
\author{Mareki \textsc{Honma}\altaffilmark{3,4,5}}
\author{Yuuki \textsc{Adachi}\altaffilmark{3}}
\author{Ross A. \textsc{Burns}\altaffilmark{1}}
\author{James O. \textsc{Chibueze}\altaffilmark{6,7}} 
\author{Yoon Kyung \textsc{Choi}\altaffilmark{8}}
\author{Kazuya \textsc{Hachisuka}\altaffilmark{3}}
\author{Kazuhiro \textsc{Hada}\altaffilmark{3,4}}
\author{Yoshiaki  \textsc{Hagiwara}\altaffilmark{9}}
\author{Shota \textsc{Hamada}\altaffilmark{10}}
\author{Toshihiro \textsc{Handa}\altaffilmark{10,11}}
\author{Mao \textsc{Hashimoto}\altaffilmark{12}}
\author{Ken \textsc{Hirano}\altaffilmark{3}}
\author{Yushi \textsc{Hirata}\altaffilmark{10}}
\author{Takanori \textsc{Ichikawa}\altaffilmark{10}}
\author{Hiroshi \textsc{Imai}\altaffilmark{10,11,13}}
\author{Daichi \textsc{Inenaga}\altaffilmark{12}}
\author{Toshio \textsc{Ishikawa}\altaffilmark{3}}
\author{Takaaki \textsc{Jike}\altaffilmark{3,4}}
\author{Osamu \textsc{Kameya}\altaffilmark{3,4}}
\author{Daichi \textsc{Kaseda}\altaffilmark{10}}
\author{Jeong Sook \textsc{Kim}\altaffilmark{14}}
\author{Jungha \textsc{Kim}\altaffilmark{2}}
\author{Mi Kyoung \textsc{Kim}\altaffilmark{3}}
\author{Hideyuki \textsc{Kobayashi}\altaffilmark{1,2}}
\author{Yusuke \textsc{Kono}\altaffilmark{1,2}}
\author{Tomoharu \textsc{Kurayama}\altaffilmark{15}}
\author{Masako \textsc{Matsuno}\altaffilmark{10}}
\author{Atsushi \textsc{Morita}\altaffilmark{10}}
\author{Kazuhito  \textsc{Motogi}\altaffilmark{16}}
\author{Takeru \textsc{Murase}\altaffilmark{10}}
\author{Akiharu \textsc{Nakagawa}\altaffilmark{10,11}}
\author{Hiroyuki \textsc{Nakanishi}\altaffilmark{10,11}}
\author{Kotaro  \textsc{Niinuma}\altaffilmark{16}}
\author{Junya \textsc{Nishi}\altaffilmark{12}}
\author{Chung Sik \textsc{Oh}\altaffilmark{14}} 
\author{Toshihiro \textsc{Omodaka}\altaffilmark{10}}
\author{Miyako \textsc{Oyadomari}\altaffilmark{13}}
\author{Tomoaki \textsc{Oyama}\altaffilmark{3}}
\author{Daisuke \textsc{Sakai}\altaffilmark{3}}
\author{Nobuyuki \textsc{Sakai}\altaffilmark{14}}
\author{Satoko  \textsc{Sawada-Satoh}\altaffilmark{16}}
\author{Katsunori M. \textsc{Shibata}\altaffilmark{1,2}}
\author{Makoto \textsc{Shizugami}\altaffilmark{3,17}} 
\author{Jumpei \textsc{Sudo}\altaffilmark{10}}
\author{Koichiro \textsc{Sugiyama}\altaffilmark{18}}
\author{Kazuyoshi \textsc{Sunada}\altaffilmark{3,4}}
\author{Syunsaku \textsc{Suzuki}\altaffilmark{1}}
\author{Ken \textsc{Takahashi}\altaffilmark{3}}
\author{Yoshiaki \textsc{Tamura}\altaffilmark{3,4}}
\author{Fumie \textsc{Tazaki}\altaffilmark{3}}
\author{Yuji \textsc{Ueno}\altaffilmark{3}}
\author{Yuri \textsc{Uno}\altaffilmark{10}}
\author{Riku \textsc{Urago}\altaffilmark{10}}
\author{Koji \textsc{Wada}\altaffilmark{10}}
\author{Yuan Wei \textsc{Wu}\altaffilmark{19}}
\author{Kazuyoshi \textsc{Yamashita}\altaffilmark{3}}
\author{Yuto \textsc{Yamashita}\altaffilmark{10}}
\author{Aya \textsc{Yamauchi}\altaffilmark{3}}
\author{Akito \textsc{Yuda}\altaffilmark{10}}
\altaffiltext{1}{Mizusawa VLBI Observatory, National Astronomical Observatory of Japan, Osawa 2-21-1, Mitaka-shi, Tokyo 181-8588}
\altaffiltext{2}{Department of Astronomical Sciences, SOKENDAI (The Graduate University for Advanced Studies), Osawa 2-21-1, Mitaka-shi, Tokyo 181-8588}
\altaffiltext{3}{Mizusawa VLBI Observatory, National Astronomical Observatory of Japan, Hoshigaoka 2-12, Mizusawa, Oshu-shi, Iwate 023-0861}
\altaffiltext{4}{Department of Astronomical Sciences, SOKENDAI (The Graduate University for Advanced Studies), Hoshigaoka 2-12, Mizusawa, Oshu-shi, Iwate 023-0861}
\altaffiltext{5}{Department of Astronomy, Graduate School of Science, The University of Tokyo, Hongo 7-3-1, Bunkyo-ku, Tokyo 113-0033}
\altaffiltext{6}{Centre for Space Research, Physics Department, North-West University, Potchefstroom 2520, South Africa}
\altaffiltext{7}{Department of Physics and Astronomy, Faculty of Physical Sciences, University of Nigeria, Carver Building, 1 University Road, Nsukka, Nigeria}
\altaffiltext{8}{Max-Planck-Institut f\"ur Radioastronomy, Auf dem H\"ugel 69, D-53121-Bonn, Germany}
\altaffiltext{9}{Toyo University, Hakusan 5-28-20, Bunkyo-ku, Tokyo 112-8606}
\altaffiltext{10}{Graduate School of Science and Engineering, Kagoshima University, Korimoto 1-21-35, Kagoshima-shi, Kagoshima 890-0065}
\altaffiltext{11}{Amanogawa Galaxy Astronomy Research Center, Graduate School of Science and Engineering, Kagoshima University,  1-21-35 Korimoto, Kagoshima 890-0065}
\altaffiltext{12}{Faculty of Science, Kagoshima University, Korimoto 1-21-35, Kagoshima-shi, Kagoshima 890-0065}
\altaffiltext{13}{Center for General Education, Institute for Comprehensive Education, Kagoshima University, 1-21-30 Korimoto, Kagoshima 890-0065}
\altaffiltext{14}{Korea Astronomy and Space Science Institute, Hwaam-dong 61-1, Yuseong-gu, Daejeon, 305-348, Republic of Korea}
\altaffiltext{15}{Teikyo University of Science, 2-2-1 Senju-Sakuragi, Adachi-ku, Tokyo 120-0045}
\altaffiltext{16}{Graduate School of Sciences and Technology for Innovation, Yamaguchi University, Yoshida 1677-1, Yamaguchi, Yamaguchi 753-8512}
\altaffiltext{17}{NAOJ ALMA Project, National Astronomical Observatory of Japan, Osawa 2-21-1, Mitaka-shi, Tokyo 181-8588}
\altaffiltext{18}{National Astronomical Research Institute of Thailand, 260 Moo 4, T. Donkaew, Amphur Maerim, Chiang Mai, 50180, Thailand}
\altaffiltext{19}{National Time Service Center, Key Laboratory of Precise Positioning and Timing Technology, Chinese Academy of Sciences, Xi'an 710600, People's Republic of China}


\maketitle

\begin{abstract}
We present the first astrometry catalog from the Japanese VLBI (very long baseline interferometer) project VERA (VLBI Exploration of Radio Astrometry). 
We have compiled all the astrometry results from VERA, providing accurate trigonometric annual parallax and proper motion measurements. 
In total, 99 maser sources are listed in the VERA catalog. 
Among them, 21 maser sources are newly reported while the rest of 78 sources are referred to previously published results or those in preparation for forthcoming papers. 
The accuracy in the VERA astrometry are revisited and compared with those from the other VLBI astrometry projects  such as BeSSeL (The Bar and Spiral Structure Legacy) Survey and GOBELINS (the Gould's Belt Distances Survey) with the VLBA (Very Long Baseline Array). 
We have confirmed that most of the astrometry results are consistent with each other, and the largest error sources are due to source structure of the maser features and their rapid variation, along with the systematic calibration errors and different analysis methods. 
Combined with the BeSSeL results, we estimate the up-to-date fundamental Galactic parameter of $R_{0}=7.92\pm0.16_{\rm{stat.}}\pm0.3_{\rm{sys.}}$~kpc and 
$\Omega_{\odot}=30.17\pm0.27_{\rm{stat.}}\pm0.3_{\rm{sys.}}$~km~s$^{-1}$~kpc$^{-1}$, where $R_{0}$ and $\Omega_{\odot}$ are the distance from the Sun to the Galactic center and the Sun's angular velocity of the Galactic circular rotation, respectively. 
\end{abstract}

\section{Introduction}

A distance toward astronomical object is the most fundamental parameter in astronomy and astrophysics. 
All the physical and dynamical properties of the target sources are estimated based on their distances. 
The most accurate and reliable method for distance determination is the trigonometric annual parallax measurements. 
For this purpose, the Hipparcos \citep{Kovalevsky1998} and the very recent GAIA data release 2 (DR2) \citep{Gaia2018} have presented optical astrometry database for more than 10$^{5}$ and 10$^{9}$ parallaxes for visible stars with accuracy of 1~mas and a few 10~$\mu$as, respectively. 
The large number of optical astrometry data play important roles for understanding not only basic properties of each target source but also statistics of various kind of stellar samples such as Hertzsprung-Russell (HR) Diagram, period-luminosity relation of variable stars, and dynamics of our Milky Way Galaxy. 
However, it is not very easy to access distant sources in the Galactic disk through observations at optical bands because of the extremely high optical extinction. 
In particular, it is crucial to determine the source distances toward dust/molecular clouds which are seen in the optical and sometimes infrared dark clouds. 

To overcome the above issues, radio astrometry observations have been developed in the last two decades by utilizing the very long baseline interferometer (VLBI) technique \citep{Reid2014a}. 
When the VLBI array consists of 1000~km baselines, the synthesized beam size (full-width half-maximum, FWHM) of an order of 1~milli-arcsecond (mas) is achievable, as roughly evaluated by $\theta_{\rm{mas}} \sim 2000\times \lambda_{\rm{cm}}/D_{\rm{km}}$ where $\lambda_{\rm{cm}}$ and $D_{\rm{km}}$ are the observed wavelengths in centimeter and the maximum baseline length in kilometer, respectively. 
Furthermore, high signal-to-noise ratios ($SNR$) allow us to measure more accurate positions of target sources than the beam size with the random (or thermal) error of $\sim \theta_{\rm{mas}}/2 SNR$ \citep{Reid1988}. 
Usually, the absolute position error in the VLBI astrometry is dominated by the systematic calibration error expressed as $\sim c\Delta\tau \theta_{\rm{SA}}/D$, where $c$, $\Delta\tau$, and $\theta_{\rm{SA}}$ are the speed of light, residual delay in calibration, and separation angle between the target source and calibrator \citep{Reid2014a}. 
If careful phase calibration are successfully conducted, 
the high accuracy of VLBI astrometry yield the trigonometric parallax for the 10~kpc (corresponding to 0.1~mas parallax) sources in the Galaxy \citep{Sanna2017, Nagayama2020a}. 

For this purpose, the VLBI Exploration of Radio Astrometry (VERA) project has been initiated in early 2000 by National Astronomical Observatory of Japan (NAOJ). 
VERA is designed to dedicate for the VLBI astrometry observations to reveal 3-dimensional velocity and spatial structures in the Galaxy. 
The observational targets are mostly limited to strong maser sources distributed across the Galaxy, with nearby position (phase) calibrators. 
The construction was completed in 2002 followed by scientific verification observations to establish method for accurate calibration and astrometry data analysis. 
Regular operations of VERA observations were started in 2004 and the first astrometry results were published in 2007 \citep{Honma2007, Hirota2007}. 
Until the end of 2019, more than 60~papers have been published to report results of VERA astrometry observations of Galactic maser sources associated with young stellar objects (YSOs) and late-type stars mostly in the asymptotic giant branch (AGB) and red supergiant (RSG) stars, as summarized in Table \ref{tab:vera}. 

At almost the same time, the Very Long Baseline Array (VLBA) legacy program named The Bar and Spiral Structure Legacy (BeSSeL) Survey has also been carrying out intensive VLBI astrometry observations mainly for distant Galactic high-mass star-forming regions (SFRs) associated with masers \citep{Reid2009a}. 
Other VLBI arrays such as the European VLBI Network (EVN) and the Australian Long Baseline Array (LBA) are also conduncting 
VLBI astrometry for maser sources in high-mass SFRs by using the 6.7~GHz CH$_{3}$OH masers \citep{Rygl2010, Krishnan2015, Krishnan2017}. 
As for low-mass nearby SFRs, another VLBA legacy survey, 
Gould's Belt Distances Survey (GOBELINS), observes non-thermal radio emission from T-Tauri stars to measure their trigonometric annual parallaxes \citep{Loinard2008, Dzib2016, Kounkel2017, Ortiz-Leon2017, Ortiz-Leon2018a}. 
Along with these surveys, number of VLBI astrometry observations have been applied to various population of stellar radio emissions \citep{Reid2014a}. 

The primary aim of this paper is to compile all the published astrometry results from VERA to establish the first VERA catalog (Section \ref{sec:results}). 
We also report some of new astrometry results from VERA in the present paper. 
The results will be compared with those of the other VLBI astrometry projects, BeSSeL (Section \ref{sec:veravlba}) and GOBELINS (Section \ref{sec:lowmass}). 
Based on the latest VERA astrometry dataset, we will revisit their accuracy and possible error sources of astrometry (Section \ref{sec:accuracy}). 
Thus, we mainly concentrate on the maser astrometry data for the samples of YSOs in SFRs and AGBs/RSGs in the present paper. 
The up-to-date Galactic constants will be estimated using all the available VLBI astrometry data (Section \ref{sec:galaxy}) based on the similar method discussed previously \citep{Reid2009b, Honma2012, Reid2014b, Honma2015, Reid2019}.

\section{Observations}
\label{sec:obs}

Although details are described in each paper, here we summarize general information about VERA astrometry observations. 
VERA is consisted of four 20~m radio telescopes in Japan; at Mizusawa, Iriki, Ogasawara, and Ishigaki-jima stations. 
The baseline lengths of VERA ranging from 1020 to 2270~km provide the synthesized beam size of 1.2~mas and 0.7~mas at 22~GHz and 43~GHz, respectively. 
We have mainly carried out astrometry observations using the $6_{1,6}$-$5_{2,3}$ transition of H$_{2}$O at 22.235080~GHz \citep{Pickett1998}. 
Some of the observations have been conducted for the $J$=1-0 SiO maser transitions at 43.122075~GHz and 42.820586~GHz for $v$=1 or/and $v$=2, respectively \citep{Muller2001}. 
To measure trigonometric annual parallaxes of maser sources, we usually carry out monitoring observations at least for 1~year and sometimes 2 years or longer. 
In some cases, monitoring observations are interrupted due to shorter life time of the target masers. 
A typical interval of monitoring is 1-2 months depending on the variability of masers; more variable sources such as AGB stars and low-mass YSOs are observed in shorter interval than 1 month. To achieve better UV coverage, each epoch of observation lasts from horizon to horizon for about 5-10 hours depending on the source declination (i.e. maximum elevation). 
In a single horizon-horizon track, we observe one or two maser sources. 
In case of observations of two different sources, we switch the target sources every 10 minutes. 

VERA astrometry observations are carried out with the dual-beam receiving system \citep{Honma2008a}. 
A pair of target maser source and reference continuum source (calibrator) is observed with two receivers simultaneously. 
The separation angle of these two sources is limited to 0.3-2.2~degrees. 
Reference sources are mainly selected from the VLBA Calibrator Catalog \citep{Beasley2002}, for which absolute positions are determined with $\sim$sub-mas accuracy. 
Some of the calibrators have been newly detected by using fringe-check survey observations with VERA at 22~GHz and/or 43~GHz \citep{Petrov2007, Petrov2012}. 
In addition to reference sources, bandpass and delay calibrator(s) are observed every 60-80 minutes. 
Amplitude calibrations are done through the chopper-wheel method \citep{Ulich1976}. 
Typical system noise temperatures are 100~K and 200~K at 22~GHz and 43~GHz, respectively, under good weather condition. However, they become higher by a factor of 2 or larger ($>$200~K at 22~GHz and $>$400~K at 43~GHz) under the conditions of high humidity and temperature, in particular at the southern isolated islands, Ogasawara and Ishigaki stations, and/or in the summer season. 

VERA can configure various frequency settings and recording settings, such as spectral resolution, total bandwidths, number of intermediate frequency (IF) channels and spectral channels within each IF. 
Details of the set-up in each observation can be seen in the respective original papers. 
In most of the observations, the digital filter output provides 16 IF channels with 16~MHz bandwidth \citep{Iguchi2005}. 
Only left-handed circular polarization is received and sampled with 2-bit per second at 1~Gbps. 
Dual-polarization observation mode is under comissioning at this moment. 
One of the 16~MHz IF channels is assigned to the target maser source and the rest of 15 IF channels are assigned to the reference source. 
For masers, a spectral resolution is set to be 15.625~kHz or 31.25~kHz, corresponding to a velocity resolution of 0.21-0.42~km~s$^{-1}$ or 0.11-0.22~km~s$^{-1}$ at 22~GHz and 43~GHz, respectively. 
Data were recorded with magnetic tapes before 2015 while more recently hard disk recording system is employed. 
The newly developed system will be capable of wider-band recoding up to 16~Gbps \citep{Oyama2016}. 
Correlation processing were made with the FX hardware correlator located at NAOJ Mitaka campus until early 2015  \citep{Chikada1991}. 
From 2015, regular operation of newly developed software correlator has been started in NAOJ Mizusawa campus \citep{Oyama2016}. 
An accumulation period in the correlation process is 1 second to produce visibility data for further post-processing data analysis. 

To achieve accurate phase calibration, reference sources are required to have flux densities higher than $\sim$50-100~mJy on average (in case of 1~Gbps recording rate) to detect their fringes with the SNR of at least 5 within a coherence time of 1-2~minutes and the recording bandwidths of 240~MHz under the best weather condition. 
Target maser sources are detectable with the peak intensities of $\sim$1~Jy~beam$^{-1}$ after successful phase-referencing analysis. 

\section{Data analysis}
\label{sec:analysis}

Basic procedures for calibration and synthesis imaging are summarized in the early results from VERA 
\citep{Honma2007, Hirota2007}. 
Calibration processes and its accuracy, in particular possible error sources by atmospheric calibration, station positions, dual-beam calibration, and source structure effects are reported in separate papers \citep{Honma2008a, Honma2008b, Honma2010, Nagayama2020a}.  
Only overall characteristics of astrometric accuracy will be discussed in the present paper. 

Before the calibration, delay tracking models are re-calculated using a software based on the CALC software package developed for the geodetic VLBI observations \citep{Manabe1991, Jike2009} because a priori models employed in the correlation processing are inaccurate for astrometry. 
In the recalculations, phase-tracking center positions of the target maser sources are shifted to the actual position of maser features within $\sim$100~mas. 
In the subsequent astrometry calibrations, more accurate delay tracking are done by using the following dataset: 
Tropospheric and ionospheric delays are recalculated based on the GPS measurements and meteorological data  \citep{Honma2008b, Nagayama2020a}, the earth orientation parameters are taken from the International Earth rotation and Reference systems Service (IERS), and the antenna positions are measured through regular monthly geodetic observations with VERA at 22~GHz \citep{Jike2009, Jike2018}. 
For the dual-beam observations, path lengths between two receiving systems for masers and reference sources are calibrated by injecting artificial noise source on the dishes \citep{Honma2008a}, which is so-called "horn-on-dish method". 
Overall calibration errors are estimated to be 10-20~mm for the tropospheric zenith delay, 3-10~TECU for the ionospheric delay, 3~mm for the antenna position, and 0.1~mm for the instrumental optical path lengths between two beams, as summarized in \citet{Nagayama2020a} and references therein. 

Other calibration processes are done in a standard manner of VLBI observations using the NRAO Astronomical Image Processing System (AIPS) software package. 
Amplitude calibrations are done by the AIPS task APCAL and ACCOR, while a template method is employed by using the AIPS task ACFIT in case of problems in chopper-wheel methods. 
The instrumental delays and phase offsets among all of the IF channels are determined by the AIPS task FRING on strong calibrator sources, and residual phases are also determined by the AIPS task FRING on reference sources. 
These delay and phase calibration solutions are copied to the target maser sources by the AIPS task TACOP, and are applied to the target maser sources by the AIPS tack CLCAL. 
If the reference sources are too weak to detect fringes, so-called inverse phase-referencing are carried out in which maser sources are used for phase calibration \citep{Hirota2011, Imai2012, Burns2015}. 
Synthesis imaging and deconvolution were performed using 
the AIPS task IMAGR. 

After making images of target maser sources at all spectral channels in all observing epochs, each maser spot and feature are identified. 
A maser spot is defined as an emission component of a single velocity channel and a feature is used for a group of spots in consecutive velocity channels at position coincident with each others. 
The maser spots and features are identified by the Gaussian fitting (AIPS tasks JMFIT or SAD) with certain threshold of noise levels in a single-channel map and integrated intensity images, respectively. 

Some of the data are analyzed by using the VEDA (VEra Data Analyzer) package developed by the VERA project for our own astrometry data \citep{Honma2007, Honma2011, Niinuma2011, Chibueze2014a, Yamauchi2016}. 
More details of the VEDA package will be presented in the forthcoming paper \citep{Nagayama2020b}. 

The identified maser spots or features are used to determine their proper motions and trigonometric annual parallaxes by fitting the positional offsets in right ascension and declination as a function of time. 
The fitting parameters are the trigonometric annual parallax $\pi$, right ascension and declination offset with respect to the tracking center position at the first epoch of observation , ($\Delta \alpha \cos \delta$, $\Delta \delta$), proper motions in right ascension and declination, ($\mu_{x}$, $\mu_{y}$)$\equiv$ ($\mu_{\alpha}\cos \delta$, $\mu_{\delta}$). 
If the astrometric accuracy is significantly worse in declination compared with that in right ascension due to different path length error in the atmospheric calibration \citep{Nagayama2020a}, only the latter data are employed to determine the parallax. 
If there are multiple maser spots or features, position offsets and proper motions for all the spots/features are fitted simultaneously with the common parallax value. 

Usually, the post-fit residual of the fitting is much larger than the astrometric accuracy expected from the thermal noise. 
This can be interpreted that there could be systematic calibration errors in the fitting results. 
Thus, the errors of these best-fit parameters are estimated by adding the noise floor to all the astrometry results in order to set the reduced $\chi^{2}$ value of unity \citep{Honma2007, Reid2009a}
This additional noise floor is regarded as the total error of astrometry including both systematic and random (thermal) errors. 

In case of VERA, error estimation is different from paper to paper. 
If there are multiple masers, two different approaches have been employed: One is to derive proper motions for individual features and  common parallax for all features simultaneously by the least squares fitting. 
In this case, the fitting error is regarded as the parallax uncertainty. 
Another method is to derive parallax value for each maser feature and average all these values. 
In this case, the standard deviation of these parallaxes is regarded as an uncertainty. 
However, if the error source of astrometry is dominated by the  calibration error of atmospheric phase fluctuation, they are common for all the features and hence, averaged parallax (the latter case) would include systematic errors which are common for all parallax values. 
Thus, the formal uncertainty could underestimate the error in the averaged parallax value. 
Although astrometry analysis methods are different from source to source in Table \ref{tab:vera}, future update of the VERA catalogue will be done by using the unified methods through the new data analysis software package VEDA \citep{Nagayama2020b}. 

\section{Results}
\label{sec:results}

In Table \ref{tab:vera}, we compile all the parallax measurements that have been conducted with VERA to date. 
In total, 99 of maser sources are listed. 
We include some sources in Table \ref{tab:vera} for which only parallax values are reported (i.e. no proper motion data). 
New astrometry results from VERA are reported for 21 sources for the first time in the present paper, while the others have been or will be published  last column of Table \ref{tab:vera}. 
Most of the target sources are observed with the 22~GHz H$_{2}$O masers, and 2 sources, Orion~KL \citep{Kim2008} and R~Aqr \citep{Kamohara2010, Min2014}, are observed in the 43~GHz vibrationally excited SiO masers. 
Numbers of YSOs in SFRs and late-type stars (AGBs and RSGs) are 68 and 31, respectively. 
Some of the target sources classified as AGBs include possible candidates of post-AGB stars or young planetary nebulae, such as IRAS~18286$-$0959 \citep{Imai2013} and K3-35 \citep{Tafoya2011}. 
Since the population of RSGs are relatively small, only 2 sources, VY~CMa \citep{Choi2008} and PZ~Cas \citep{Kusuno2013} are reported. 

We refer the astrometry data from the original papers as listed in the last column in Table \ref{tab:vera}. 
Numbers of significant digits of astrometric parameters are different from source to source, depending on the relative uncertainties. 
We simply set the uniform number of significant digit for the parameters in Table \ref{tab:vera} except for the parallax values, for which we follow the definition of the original papers. 

For proper motions, we need additional consideration in some cases. 
When the target sources are associated with multiple maser features, many literature calculated their averages to estimate systemic motions which are regarded to represent those of central stars. 
On the other hand, if there are insufficient number of maser feature(s) in a target source, a proper motion of single feature is used to estimate its systematic motion. 
If there is no explicit discussion on proper motions in the original paper, we calculate these values as mentioned above. 
In most of the VERA results, errors in the proper motions are determined by the formal uncertainties in the fitting in case of sources with a single feature or the mean proper motions of multiple maser features. 
To ensure the use of an accurate estimate of the source systemic proper motion in modelling the Galactic rotation, it is necessary to separate observed proper motions into their respective contributions from the internal proper motions caused by jets, outflows etc., and the true motion of the SFRs in the sky plane. 
In cases where the proper motions are symmetric or random, this can be done by simply averaging all measured proper motions. 
However, in asymmetricaly sampled cases, or cases of few detected maser features, the average motion may misrepresent the source proper motion. 
This consequently introduces errors into the evaluation of the model parameters during fitting. 
We did not consider such potential systematic uncertainties in the proper motions in Table \ref{tab:vera}.  
However, if a large enough sample of sources is used then these errors introduced should average out. 

The radial velocities of the target sources, which are usually measured by the radio molecular lines such as CO and NH$_{3}$ or by the maser lines themselves, are also listed in Table \ref{tab:vera}. 
Similarly, these radial velocities could result in significant amount of uncertainties in the estimated 3-dimensional velocity field of the Galaxy. 
In particular, the definition of the radial velocities would affect the estimation of the systemic velocity of the target sources depending on either average or central velocities of the maser features. 
It is known that the H$_{2}$O masers show sometimes extremely high velocity features up to 10-100~km~s$^{-1}$ with respect to the systemic velocity \citep{Motogi2016}. 
For instance, one of the target sources, IRAS~20255$+$4032, shows the average velocity of the four maser features of $-63.3$~km~s$^{-1}$, while the systemic velocity is determined by the CO line to be $-18.2$~km~s$^{-1}$ \citep{Sakai2020c}. 
In such cases, maser data will lead erroneous assumption of the systemic velocities. 
If there is no estimation of uncertainties in the radial velocity, we take into account these uncertainties of 5~km~s$^{-1}$ as indicated in Table \ref{tab:vera}. 
Although the BeSSeL project employs the more conservative value of 10~km~s$^{-1}$, our results are not severely affected these different assumptions. 

\section{Discussion}
\label{sec:discussion}

In this section, we will revisit discussion on accuracy of the VERA astrometry and estimation of the Galactic fundamental parameters, as reported in previous summary papers \citep{Reid2009b, Honma2012, Reid2014a, Reid2014b, Honma2015, Reid2019}. 

\subsection{Comparison with VERA and VLBA/EVN results}
\label{sec:veravlba}

\begin{figure}
\begin{center}
\includegraphics[width=8cm]{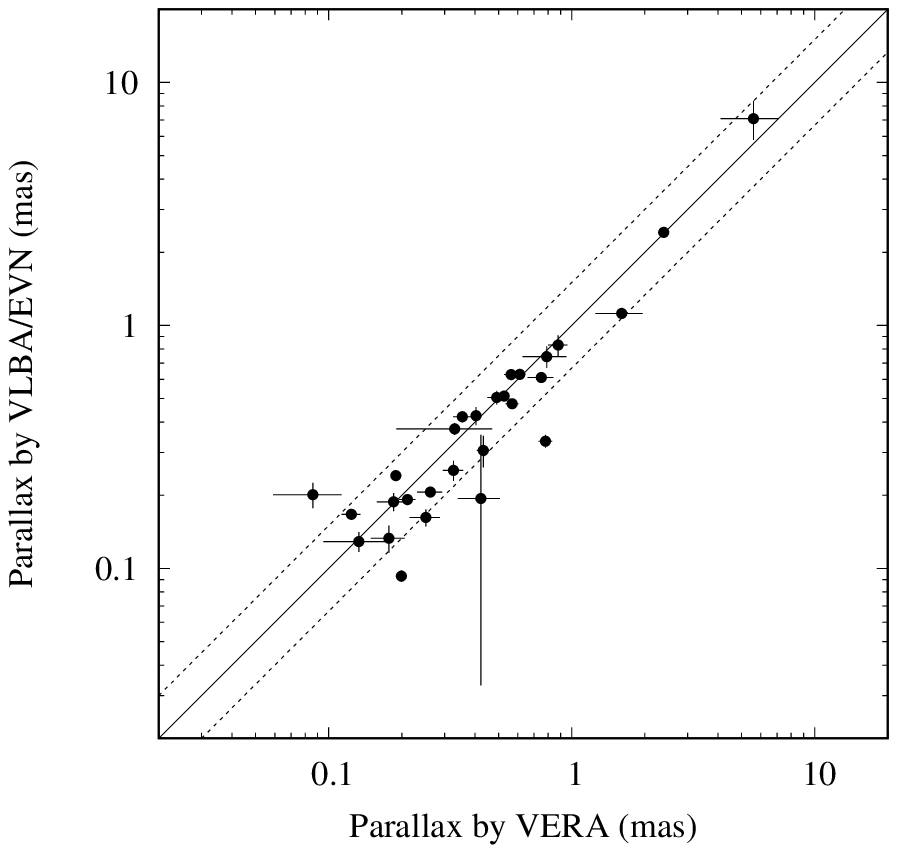}
\end{center}
\caption{Comparison of parallaxes from VERA and VLBA/EVN. 
A solid line indicates the VERA parallaxes equal to those of VLBA/EVN while dashed lines show their differences by factors of 1/1.5 and 1.5. 
}
\label{fig:VERA-VLBA}
\end{figure}

Figure \ref{fig:VERA-VLBA} compares the results of parallax measurements carried out by VERA and VLBA/EVN. 
For this plot, most of the target sources are the H$_{2}$O and/or CH$_{3}$OH maser sources which are identified in the same SFRs within 1\arcmin \ as listed in Table \ref{tab:compare}. 
Some of high-mass SFRs host multiple YSOs associated with different maser clusters within individual regions. 
We do not include such sources because they could be located in different molecular clouds aligned along the line-of-sight by chance. 
Only exceptions are Orion~KL and HH~12-15 which are observed in radio continuum emission \citep{Menten2007, Dzib2016}. 
For Orion~KL, we refer to \citet{Menten2007}, although there are multiple/different VLBI astrometry results for different YSOs in the same region (see more discussion in the next section). 
To compare these VLBI astrometry results with each other, we employ the astrometric parameters from VLBA/EVN reported in original references with the highest accuracy data for each source rather than those compiled in \citet{Reid2019} because some of their results are averaged value of multiple VLBI astrometry results. 
For G359.62$-$00.25 and Sgr~D/G001.14$-$00.12, we use the parallax values in \citet{Reid2019} because the references are in preparation. 

As seen in the clear correlation in Figure \ref{fig:VERA-VLBA}, most of the astrometry results are consistent within a factor of 1.5 except for a few sources with larger scatter. 
They are IRAS~05137$+$3919/G168.06$+$00.82 \citep{Honma2011, Hachisuka2015}, 
Sgr~D/G001.14$-$00.12 \citep{Sakai2017, Reid2019}\footnote{Although it is the largest error bar in the plot, we could not confirm the original reference in \citet{Reid2019}.}, G005.88$-$00.39 \citep{Motogi2011, Sato2014}, and G048.60$+$00.02 \citep{Nagayama2011a, Zhang2013}. 
These data show differences larger than a factor of 2. 

\begin{figure}
\begin{center}
\includegraphics[width=8cm]{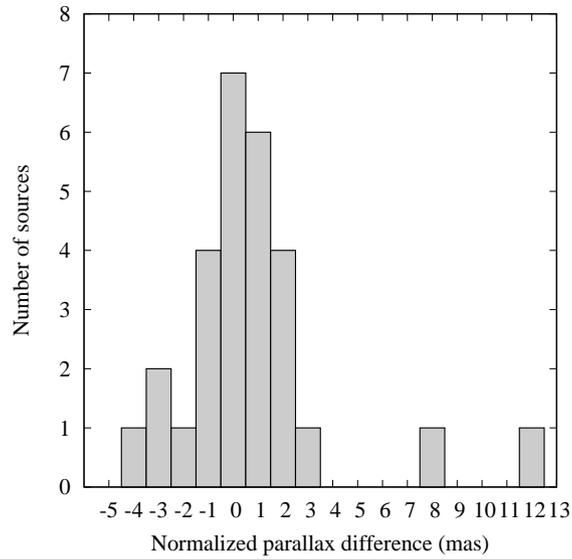}
\end{center}
\caption{Histogram of the normalized parallax differences scaled by their joint uncertainties. See definition in equation \ref{eq:norm}. 
Each bin has the central value of integer with its width of 1 (e.g. $\Delta \pi/\sigma_{\Delta \pi}=0.0\pm0.5$, $1.0\pm0.5$, $-1.0\pm0.5$, ...).  
}
\label{fig:norm}
\end{figure}

To compare their differences, we calculate the normalized parallax differences scaled by their joint uncertainties, as defined by the following equation; 
\begin{equation}
\Delta \pi/\sigma_{\Delta \pi} \equiv \frac{\pi_{\rm{VERA}}-\pi_{\rm{VLBA}}}{\sqrt{\sigma_{\rm{VERA}}^{2}+\sigma_{\rm{VLBA}}^{2}}}
\label{eq:norm}
\end{equation}
in which $\pi_{\rm{VERA}}$ and $\pi_{\rm{VLBA}}$ are the parallaxes measured with VERA and VLBA, respectively, and $\sigma_{\rm{VERA}}$ and $\sigma_{\rm{VLBA}}$ are the parallax errors for VERA and VLBA results, respectively. 
The results are listed in Table \ref{tab:compare} and the  distribution is plotted in Figure \ref{fig:norm}. 
As seen in Figure \ref{fig:norm}, most of the target sources give consistent results within $\Delta \pi/\sigma_{\Delta \pi} <3$ or less than 3$\sigma$ limit. 
The first quartile, median, and third quartile are $-0.65$, 0.35, and 1.49, respectively. 
The largest discrepant measurements are for G048.60$+$00.02 (12.32), G005.88$-$00.39 (8.28), S269 ($-3.61$), G135.28$+$02.80 ($-3.43$), and IRAS~05137$+$3919 ($-3.18$). 
Sgr~D/G001.14$-$00.12 give the smaller value of 1.26 due to the exceptionaly large relative errors of VLBA parallax of $\sim$80\%. 
We note that the possible systematic errors in the VERA parallax as discussed in the last paragraph of Section \ref{sec:results} are not included in the calculated $\Delta \pi/\sigma_{\Delta \pi}$. 
Thus, parallax measurements from VERA and VLBA/EVN for 23 out of total 28 samples (82\%) agree with each others within 3$\sigma$ levels. 

Possible origins of such large errors are due to insufficient number of observing epochs, in particular around the peak season of the annual parallax value (for IRAS~05137$+$3919), or errors in the atmospheric calibration and VERA dual-beam phase calibrations (for G048.60$+$00.02). These results could be improved by using additional data and re-calibration processing. 
Furthermore, spatially extended structures of the target maser features could significantly degrade the accuracy of the position measurements of the maser features, which would introduce additional errors in the astrometry and hence, derived parallax values. 
In fact, the possible effect of such maser structures is intensively discussed for a high-mass YSO S269 at the distance of 4~kpc \citep{Honma2007, Asaki2014, Quiroga-Nunez2019}, in which the possible structure effect results in a parallax error of $>$20\%. 
It has been already discussed for the VERA data \citep{Honma2010} and we will revisit this issue in the next sections. 

\subsection{Comparison for distances toward nearby SFRs}
\label{sec:lowmass}

Several astrometry results are reported for the Orion Molecular Cloud both with VERA and VLBA since the beginning of the VERA and VLBA astrometry projects \citep{Hirota2007, Kim2008, Sandstrom2007, Menten2007, Kounkel2017} including the new result from VERA \citep{Nagayama2020b}. 
In the central part of the Orion region, active high-mass SFRs Orion~KL and Orion Nebular Cluster (ONC) are of great interest.
The VERA results are for observations of the H$_{2}$O masers or SiO masers (Orion Source~I), while VLBA observes radio continuum emission from different non-thermal radio emitting YSOs in the ONC region. 
The first astrometry results for these sources had larger uncertainties of 437$\pm$19~pc \citep{Hirota2007} and 389$^{+24}_{-21}$~pc ~pc \citep{Sandstrom2007} from the VERA H$_{2}$O maser and VLBA 15~GHz continuum observations, respectively. 
Subsequent higher accuracy data of 418$\pm$6~pc from the SiO masers with VERA \citep{Kim2008} and 414$\pm$7~pc from the 8~GHz continuum with VLBA \citep{Menten2007} are in excellent agreement. 
These results suggest a weighted-mean distance of $416.3\pm4.6$~pc toward the Orion~KL region or the central part of the ONC. 
On the other hand, recent comprehensive studies with VLBA by \citet{Kounkel2017} suggest a smaller distances of 388$\pm$5~pc as a weighted average of distances of YSOs in wider area of ONC. 

The possible reason for the differences in these parallax measurements, in particular compared with that of \citet{Menten2007}, are discussed in \citet{Kounkel2017}, which are attributed to the different number of samples, systematic errors originated from ionospheric calibration, multiplicity of the target sources, and/or different treatment of the data for the fitting (e.g. fitting routine for the right ascension and/or declination). 
If this difference is real, it would suggest the depth of the region along the line-of-sight; Source~I could be located in the rear side of the ONC which argues against \citet{Kim2008}. 

For other nearby low-mass star-forming regions, we have carried out series of astrometry observations \citep{Imai2007, Hirota2007, Hirota2008a, Hirota2008b, Kim2008, Hirota2011}. 
Similar comprehensive survey are also carried out by the VLBA large program GOBELINS and their pilot surveys \citep{Loinard2008, Dzib2016, Kounkel2017, Ortiz-Leon2017, Ortiz-Leon2018a}. 
Some of the target regions are common, such as Orion, Monoceros, Ophiuchus, and Perseus molecular clouds. 
In Figure \ref{fig:VERA-VLBA}, only Orion~KL, HH~12-15, and IRAS~16293$-$2422 in L1689 are plotted as they are regarded as the same SFRs observed with VERA. 

We quantitatively compare the difference in distances  between two astrometry measurements. 
Here we compare distances rather than parallaxes because many literature listed mean distances of multiple sources. 
In the following discussion, the error in the distance difference is calculated from the root sum square of each distance error, and the error in each distance is estimated from the geometric mean of both error bars (i.e. $\Delta D=\sqrt{D_{1}D_{2}}$ in case of $D^{+D_{1}}_{-D_{2}}$). 
The difference between relative errors in the parallax and distance estimated above are as small $<$0.1-0.3\%. 

For the Orion regions, GOBELINS also includes various molecular clouds outside Orion~KL. 
One of examples is the L1641 region at the measured distance of 428$\pm$10~pc \citep{Kounkel2017}. 
The VERA result for another nearby maser source, L1641~S3, presents the slightly larger distance of 473$^{+32}_{-27}$~pc. 
The difference in these two distance values of 45$\pm$31~pc is not significant with the only 1.5$\sigma$ level. 
Although VLBA failed to determine the parallax of YSOs in $\lambda$~Ori possibly due to scattering at the lower frequency \citep{Kounkel2017}, we can measure the parallax of a YSO associated with the $\lambda$~Ori region, B35, to be 1.98$\pm$0.25~mas in the present paper, corresponding to the distance of 505$^{+73}_{-57}$~pc. 
The difference in distances between those from GAIA~DR2, 402$\pm$1$\pm$20~pc \citep{Zucker2019}, and VERA is 103$\pm$68~pc, which is 1.5$\sigma$ level. 
Because the uncertainty in the parallax value from VERA is still large, future more accurate observations are required to confirm the result. 

For YSOs in the Monoceros region, HH (or GGD) 12-15, distances from VERA and VLBA of 620$^{+180}_{-110}$~pc (present paper) and 893$^{+44}_{-40}$~pc \citep{Dzib2016}, respectively, are different by a factor of 45\%. 
The difference in these two distances is $-273\pm147$~pc. 
The error bar of the VERA result, $\sim$20\%, is relatively larger the typical value (see Figure \ref{fig:pi-dpi}) as discussed in the next section. 

For the Ophiuchus region, the distances measured by GOBELINS are 137.3$\pm$1.2~pc and 147.3$\pm$3.4~pc toward dark clouds L1688 and L1689, respectively \citep{Ortiz-Leon2017}. 
Parallax measurements for a protostar IRAS~16293$-$2422, which is located in L1689, give 178$^{+18}_{-37}$~pc and 141$^{+30}_{-21}$~pc from VERA \citep{Imai2007} and VLBA \citep{Dzib2018a}, respectively. 
These two values from maser astrometry marginally agree with each other (37$\pm$36~pc), but the smaller distance of $\sim140$~pc is more consistent with those from continuum sources in L1689. 

Comparing with the GAIA~DR2 results, VLBA results are confirmed to be in good agreement for Ophiuchus \citep{Ortiz-Leon2017, Ortiz-Leon2018b} and Perseus \citep{Ortiz-Leon2018a} regions. 
In case of another nearby SFR, NGC2264, the distance measured with VERA of 738$^{+57}_{-50}$~pc \citep{Kamezaki2014a} is in good agreement with the GAIA DR2 value, 719$\pm$16~pc \citep{MaizApellaniz2019}. 
The mean distance toward slightly farther molecular clouds in Gem~OB1, IRAS~06058$+$2138 \citep{Oh2010}, IRAS~06061$+$2151 \citep{Niinuma2011}, and S255~IR-SMA1 \citep{Burns2016}, of 1.85~kpc (with a standard deviation of 0.14~kpc) is also consistent with that from the GAIA DR2 result, 1.786$\pm$0.004$\pm$0.089~kpc \citep{Zucker2019}. 
On the other hand, some of the parallax values derived from VERA observations showed significantly large uncertainties which are larger differences than the error bars of VLBA and GAIA results.  
In the case of Perseus Molecular Cloud, the VERA results of 234$\pm$13~pc from the weighted mean distance of NGC~1333 \citep{Hirota2008a} and L1448 \citep{Hirota2011} is smaller by $-59\pm26$~pc than that of GAIA DR2 of 293$\pm$22~pc  \citet{Ortiz-Leon2018a}, although the parallax of NGC~1333 was not determined with VLBA alone. 

We note that the parallax values of GAIA DR2 would include the zero-point offset with an order of $-0.1$-0~mas \citep{Gaia2018}. 
More detailed analysis are presented in \citet{Xu2019} and references therein, suggesting that the parallax offset in the GAIA DR2 data is $-75\pm29$~$\mu$as. 
It requires the correction of the parallax value corresponding to the systematic uncertainty from $-1.5$~pc to $-180$~pc at the distance of 140-1800~pc in the SFRs discussed above. 
This effect is already considered as the possible systematic error in each reference. 
In addition, the parallax offset in the GAIA DR2 data would affect more significant in the farther target sources. 
Thus, the large differences in the parallax values for nearby SFRs are mostly due to shorter lifetime of the H$_{2}$O masers associated with the low-mass YSOs than the period of annual parallax (i.e. 1~year), and there seems no significant contribution from the zero-point offset in the GAIA DR2 parallaxes. 
In addition, significant spatial structures of the nearby sources could degrade the accuracy of the position measurements \citep{Imai2007, Hirota2008b, Honma2010, Dzib2018a}. 
We will evaluate this effect in the next section. 

\subsection{Accuracy and dominant error sources in VERA Astrometry}
\label{sec:accuracy}

As discussed in \citet{Honma2010}, a motion of 0.5~km~s$^{-1}$, which is comparable to typical line widths of the masers (full-width half-maximum of 1~km~s$^{-1}$), corresponds to the transverse distance of 0.1~au within 1~yr. 
If the motion is systematic, it can partly contribute to the linear proper motion and hence, is measurable by the VLBI astrometry monitoring. 
However, if such a motion is originated from a turbulence in the maser cloud, it could cause change in the structure of the maser feature randomly. 
The possible structure change would affect the positional accuracy of the maser features. 
As a result, this effect will lead to the error in the annual parallax, which is equal to the angular size of 1~au at the distance of the target source, up to 10\%.  
Even larger errors up to 20\% or corresponding structure changes of 0.2~au scale would be likely, given the spatially extended nature of maser features ($>$1~au). 

\begin{figure}
\begin{center}
\includegraphics[width=8cm]{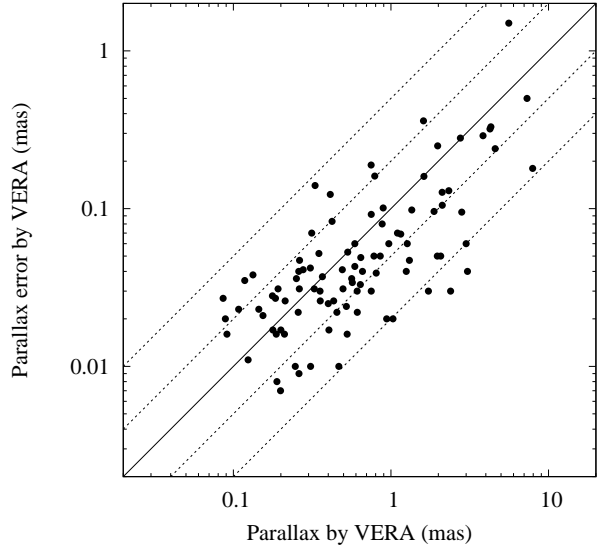}
\end{center}
\caption{Parallax values and their errors obtained from the VERA astrometry. 
A solid line indicates the parallax errors of 10\%, while dashed lines represent the errors of 50\%, 20\%, 5\%, and 2\%. }
\label{fig:pi-dpi}
\end{figure}

Figure \ref{fig:pi-dpi} shows the errors in the parallax as a function of the parallax values. 
Obviously, the plot shows a clear trend of correlation as seen in the smaller number of samples \citep{Honma2010}. 
In other word, the errors in the parallaxes are mostly 2-20\% independent of the source distances. 
It is unlikely that the correlation is mostly due to the calibration errors as they should strongly depend on the separation angles between the calibrators and targets, weather condition, and source elevation (declination), rather than their distances. 

It should be noted that the larger variation in Figure \ref{fig:pi-dpi} than that of \citet{Honma2010} would also reflect different calibration errors and/or method of analysis such as different number of maser features employed in the parallax fitting and lengths of astrometry monitoring observations. 
In general, lower declination sources at $\delta<-30$~degrees, such as NGC6334(I), are more seriously affected by atmospheric calibration errors \citep{Chibueze2014a}. 
However, we confirm that the correlation of parallaxes and their errors would be the results of source structures. 

For the AGB stars, trigonometric parallax measurements can be done by using the VLBI astrometry of maser sources and optical astrometry like Hipparcos and GAIA (DR2). 
One example for a semi-regular variable star SV~Peg demonstrates that there could be significant differences between VLBI and GAIA DR2 astrometry, for which parallax values are 3.00$\pm$0.06~mas and 1.12$\pm$0.28~mas, respectively \citep{Sudou2019}. 
The discrepancy is most likely attributed to an effect of unresolved structure of the stellar photosphere observed with GAIA DR2.   
\citet{Xu2019} discussed accuracy of parallax measurements of YSOs, AGBs, and pulsars from VLBI astrometry and GAIA DR2, and found the largest differences in the AGB samples. 
More detailed comparison of astrometry observations for each AGB star will be discussed in a separate paper \citep{Matsuno2020, Nakagawa2020} and hence, it is out of the scope of this paper. 

The structure effect is though to be more serious for highly variable sources such as low-mass nearby YSOs and AGB stars \citep{Imai2007, Hirota2008b}. 
In contrast, it is demonstrated that astrometry for the compact stellar emission could achieve $<$1\% parallax accuracy for nearby open cluster Pleiades \citep{Melis2014} and continuum sources in the nearby Ophiuchus molecular clouds \citep{Ortiz-Leon2017}. 
The structure changes in maser features are found to be less significant for farther ($\sim$10~kpc) sources because the other error budgets, in particular due to calibration errors of tropospheric delay term \citep{Honma2008b, Nagayama2020a}, become more significant than those from the source structure.
Hence, we note that target maser sources should be selected carefully according to their spatial structures in the synthesized images to overcome this issue. 
It is also important to make images with better uv coverages to recover both spatially compact and extended emission components. 
In case of VERA, lack of shorter baselines ($<$1000~km) would seriously resolve out spatially extended maser features. 
Thus, further KaVA (KVN and VERA Array) and EAVN (East Asian VLBI Network; An et al. 2018) are expected to improve the accuracy of the maser astrometry. 

\subsection{Galactic Structure}
\label{sec:galaxy}

\begin{figure}
\begin{center}
\includegraphics[width=8cm]{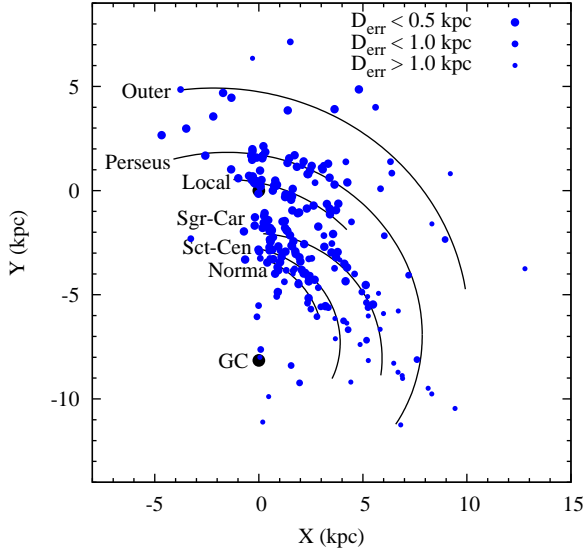}
\end{center}
\caption{Distributions of the maser sources on the face-on view of the Galaxy.
Solid line show the spiral arm structure identified by the BeSSeL results \citep{Reid2019}. }
\label{fig:faceon}
\end{figure}

The currently available VLBI astrometry results are plotted in Figure \ref{fig:faceon}. 
We plot positions of SFRs and RSGs but AGBs are not included because these sources will not be used for the further analysis as discussion later. 
Thus, total 224 sources are selected including both VERA and other VLBI results \citep{Reid2019}. 
We indicate the location of the best-fitted Galactic spiral arms determined by \citet{Reid2019}. 
Most of the target sources are located in the northern hemisphere with the declination of $\delta>-35$~degrees because of the visibility of the target sources from VERA, VLBA, and EVN. 
Thus, the sample distribution is strongly biased to the first and second Galactic quadrants ($0$~deg$<l<180$~deg, where $l$ is the Galactic longitude), while less number of sources are located in the third quadrant ($180$~deg$<l<270$~deg). 
Only exceptions are two sources, G339.884$-$1.259 \citep{Krishnan2015} and G305.2 region \citep{Krishnan2017}, which are observed with the Australian LBA, as plotted in the fourth quadrant ($270$~deg$<l<360$~deg) of Figure \ref{fig:faceon}. 
The most distant parallax measurement with VLBI is achieved for a high-mass SFR G007.47$+$00.05 with the trigonometric parallax from the BeSSeL project of 0.049$\pm$0.006~mas, corresponding to 20.4$^{+2.8}_{-2.2}$~kpc \citep{Sanna2017}. 
This value is consistent with the astrometry observations with VERA of 20$\pm$2~kpc, which is estimated based on the absolute proper motions and radial velocity measurements and  3-dimensional Galactic rotation model \citep{Yamauchi2016}. 

\begin{figure}
\begin{center}
\includegraphics[width=8cm]{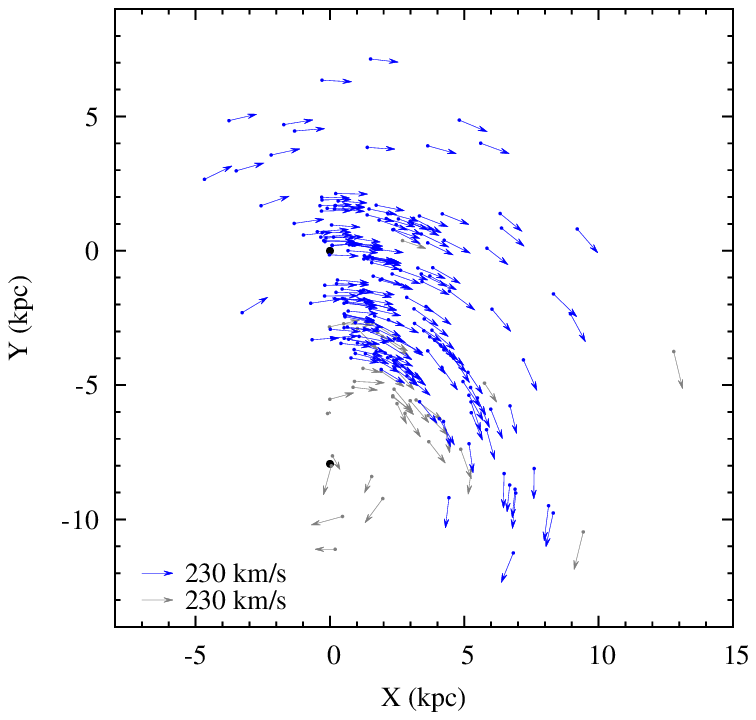}
\includegraphics[width=8cm]{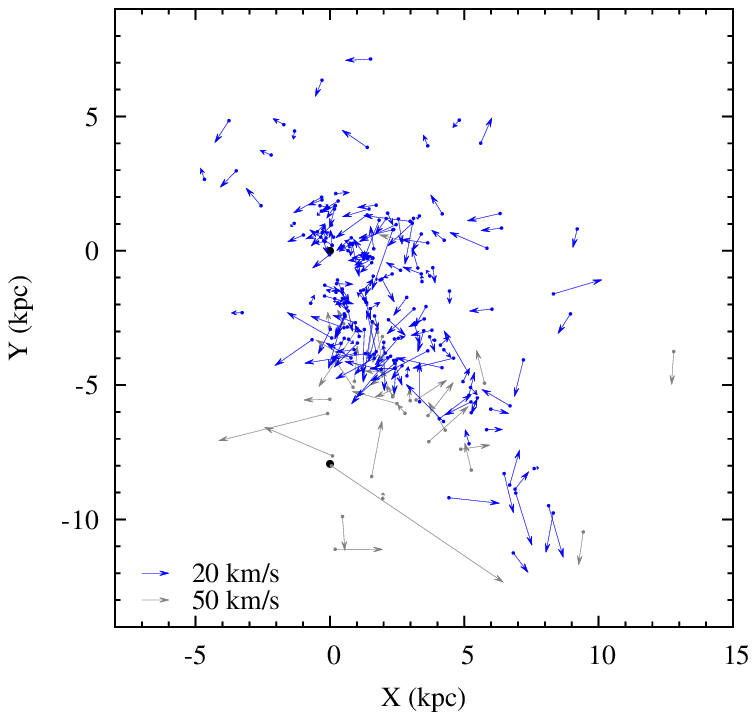}
\end{center}
\caption{Galactic rotation motions (left) and peculiar motions (right) of the maser sources. 
Gray symbols indicate the outliers with $R< 4$~kpc or the peculiar motion of $V>$50~km~s$^{-1}$, which are removed from the MCMC analysis (see text). 
Rest of the sources are plotted in the blue symbols. 
Number of blue and gray symbols are 189 and 35, respectively (total 224 sources). 
The Galactic parameters for the model with the power-law rotation curve (Table \ref{tab:prm}) and the Solar motion of ($U_{\odot}$, $V_{\odot}$, $W_{\odot}$) = (11.1, 12.24, 7.25) in km~s~$^{-1}$ \citep{Schonrich2010} are employed to plot the vectors. }
\label{fig:vector}
\end{figure}

\begin{figure}
\begin{center}
\includegraphics[width=8cm]{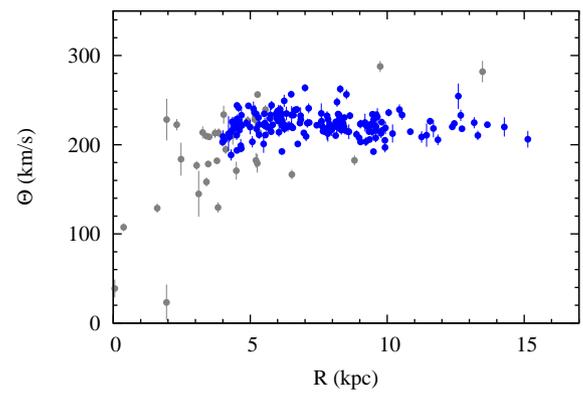}
\end{center}
\caption{Rotation curve of the Galaxy.
The blue and gray symbols are the same as in Figure \ref{fig:vector}. }
\label{fig:rotation}
\end{figure}

The Galactic rotation can be seen in Figure \ref{fig:vector} in which positions of maser sources are plotted with 2-dimensional velocity vectors in the Galactic plane. 
The 2-dimensional vectors in the Galactic plane projection are determined by combination of sky plane and line of sight velocities as mentioned in Section \ref{sec:results}. 
To transform from the measurements in a Heliocentric frame to the Galacto-centric reference frame, we employ the Galaxy model with the power-law rotation curve as discussed below. 
The parameters used for the transformation are summarized in Table \ref{tab:prm}. 
We also assume the Solar motion of ($U_{\odot}$, $V_{\odot}$, $W_{\odot}$) = (11.1, 12.24, 7.25) in km~s~$^{-1}$ \citep{Schonrich2010}. 
The Galactic rotation curve can be constructed as plotted in Figure \ref{fig:rotation} using the same parameter set. 
The well known flat rotation curve is confirmed toward the  distance up to 15~kpc from the Galactic center. 

By combining currently available maser astrometry results from VERA, VLBA, EVN, and LBA, we can estimate the fundamental parameters of the Milky Way Galaxy as discussed in \citet{Honma2012} by using increased number of samples. 
Here we briefly summarize our data analysis. 
We employ SFRs and high-mass RSGs but exclude AGB stars for our data analysis. 
This is because dynamical properties of AGB stars are different from those of former samples, such as velocity dispersion and peculiar motions with respect to the Galactic rotation (known as asymmetric drift). 
In the model fitting, outliers which have the Galacto-centric distances within $<4$~kpc or the peculiar motion of  $V>$50~km~s$^{-1}$ are also removed from the input data for further analysis to avoid systematic errors in the estimated parameters. 
The former condition is considered to exclude systematic motion caused by the Galactic bar \citep{Honma2012, Reid2014b}. 
These sources are indicated by gray symbols in Figures \ref{fig:vector} and \ref{fig:rotation}. 
The number of removed sources is 35, and hence, we used total 189 sources for the further analysis, which are plotted by blue symbols in Figures \ref{fig:vector} and \ref{fig:rotation}. 

In the Galaxy model, we simply assume the circular rotation of the Local Standard of Rest (LSR) with small systematic/random motions. 
The distances toward the Galactic center and the rotation velocity of the LSR around the Galactic center are referred to $R_{0}$ and $\Theta_{0}$, respectively.
The ratio of $\Theta_{0}/R_{0}$ gives the Galactic angular velocity at the LSR, $\Omega_{0}$. 
As discussed previously, we will solve a set of $R_{0}$ and $\Omega_{0}$, rather than $R_{0}$ and $\Theta_{0}$ because the latter set is known to be tightly correlated \citep{Reid2009b, Honma2012}. 
Although the correlation could become modest due to increased number of target sources in larger distribution of our present data, we follow the same data analysis by \citet{Honma2012} to compare the results consistently. 
For the Galactic rotation curve, we use two different models; a power law, $\Theta(R) = \Theta_{0} (R/R_{0})^{\alpha}$, and 2nd-order polynomial, 
$\Theta(R) = \Theta_{0} + a_{0} (R-R_{0}) + b_{0} (R-R_{0})^{2}$, functions of rotation curves. 
The power-law index $\alpha$, or the polynomial coefficients $a_{0}$ and $b_{0}$ are also solved in the analysis. 
In addition, mean systematic motions, $(U_{\rm{s}}, V_{\rm{s}}, W_{\rm{s}})$ are introduced in the models to account for the peculiar motions \citep{Reid2009b, Honma2012}. 
The $U_{\rm{s}}$, $V_{\rm{s}}$, and $W_{\rm{s}}$ are defined as the velocity components toward the Galaxy center, the direction of Galactic rotation, and the north Galactic pole, respectively.
The Solar motion is assumed to be ($U_{\odot}$, $V_{\odot}$, $W_{\odot}$) = (11.1, 12.24, 7.25) in km~s~$^{-1}$ \citep{Schonrich2010}. 

\begin{figure*}
\begin{center}
\includegraphics[width=16cm]{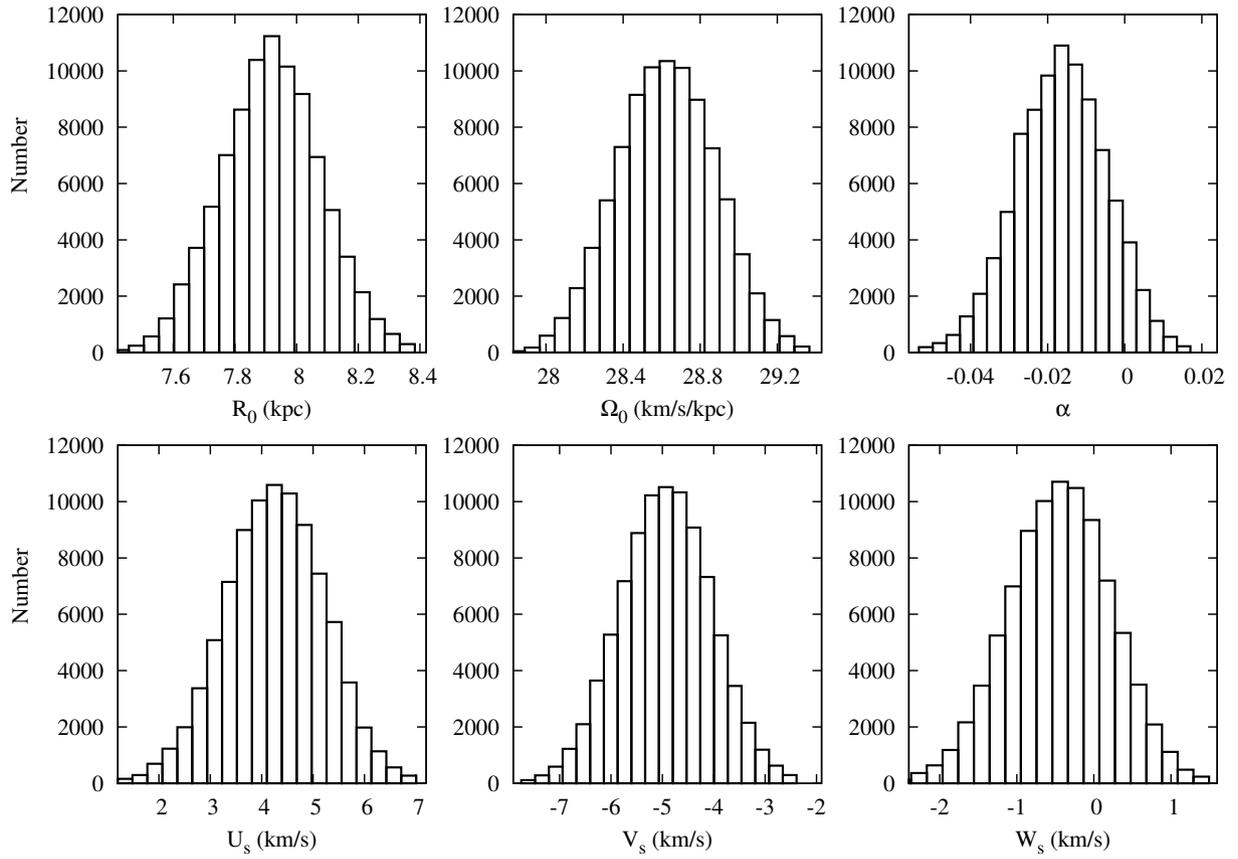}
\end{center}
\caption{Posterior provability distribution for the Galactic parameters in the case of the power law model. 
From top-left to bottom-right, each panel shows a plot for $R_{0}$, $\Omega_{0}$, $\alpha$, $U_{\rm{s}}$, $V_{\rm{s}}$, and $W_{\rm{s}}$. }
\label{fig:hist}
\end{figure*}

All the parameters are estimated by the same procedures described in \citet{Honma2012}, based on the Markov Chain Monte Carlo (MCMC) method. 
To explore the posterior provability distribution of the parameters, MCMC simulation is iterated for the trial number of $10^6$. 
Figure \ref{fig:hist} shows the posterior probability distribution for six parameters of power-law model. 
For all parameters, the posterior profitability distribution shows a single-peaked symmetric structure, which confirms reasonable estimates of the parameters. 
Table \ref{tab:prm} summarize the best parameters and their statistical errors calculated from the means and the standard deviations of the posterior probability distributions in our two Galactic rotation models. 
As for $R_{0}$ and $\Omega_{0}$, both results agree well with each other with differences less than 1\%. 
These differences are much smaller than the error bars. 
The inward motion of $U_{\rm{s}}$ is non-zero values in contrast to the previous paper \citep{Honma2012}, while the vertical component with respect to the Galactic plane, $W_{\rm{s}}$ suggests no significant motion. 
The power-law index $\alpha$ of $-0.016\pm0.012$ is slightly negative but is consistent with the flat rotation curve. 

As listed in Table \ref{tab:prm}, we can directly compare the present results with the models of ID~14 for the power-law rotation curve and ID~22 for the polynomial rotation curve (removing outliers and adopting fixed $V_{\odot}=12$~km~s$^{-1}$) in \citet{Honma2012}. 
Present results are in good agreement with those in \citet{Honma2012} but the statistical errors for $R_{0}$ and $\Omega_{0}$ estimated by the MCMC method are smaller by a factor of $\sim$2.5 as the number of samples increases from 49 to 189. 
In particular, an increase in the number of further sources would significantly contribute to improve the precision. 

As reported in the previous paper \citep{Honma2012}, $\Omega_{0}$ and $V_{\odot}$ is tightly correlated and $\Omega_{0}$ is dependent on the adopted value of $V_{\odot}$. However, the angular velocity of the Sun defined by $\Omega_{\odot} \equiv \Omega_{0}+V_{\odot}/R_{0}$ can be well determined. 
It is estimated to be $\Omega_{\odot}$=30.17$\pm$0.27~km~s$^{-1}$~kpc$^{-1}$ using $R_{0}$ and $\Omega_{0}$ in the power-law model shown in Table \ref{tab:prm} and the adopted value of $V_{\odot}$=12.24~km~s$^{-1}$ \citep{Schonrich2010}. 

As demonstrated in \citet{Honma2012}, estimated Galactic fundamental parameters depend on the employed model of the Galaxy and input source samples. 
In the present paper, we did not perform the MCMC fitting by changing the Solar motion. 
The assumed values of the Solar motion would lead to systematic errors in the derived parameters and hence, the uncertainties in the estimated parameters could be underestimated. 
For instance, the difference in $R_{0}$ is as small as $<1$\% while that of $\Omega_{0}$ is about 6\% when $V_{\odot}$ is assumed to be 5.25~km~s$^{-1}$ or 12.0~km~s$^{-1}$, or is solved in the MCMC analysis to be $\sim$19~km~s$^{-1}$ (Table 5 and Figure 4 in Honma et al. 2012). 
The different values of the Solar motion also result in the systematic error of $\Theta_{0}$ of the same magnitude, 6\% (Table 6 in Honma et al. 2012). 
However, it does not strongly affect the $\Omega_{\odot}$ value, which is 0.3\% difference (Table 6 in Honma et al. 2012). 
In summary, assuming the Solar motions in the MCMC analysis would lead systematic errors of 6\% in $\Omega_{0}$ and $\Theta_{0}$, while the effect is less than 1\% for $R_{0}$ and $\Omega_{\odot}$. 
In order to take into account the above systematic errors, we estimate the systematic error in $R_{0}$ of 4\%, adding the 1\% of the model dependency (Table \ref{tab:prm}) and 3\% of the sample dependency \citep{Honma2015}. 
For the $\Omega_{\odot}$, the systematic error is 1\% mainly due to the sample dependency \citep{Honma2015}, given the smaller differences among models in Table \ref{tab:prm}  \citep{Honma2012, Honma2015, Reid2019}. 

According to \citet{Reid2019}, $R_{0}$ and $\Omega_{\odot}$ are determined to be 8.15$\pm$0.15~kpc and $30.32\pm0.27$~km~s$^{-1}$~kpc$^{-1}$, respectively, based on the VLBI astrometry of 147 maser sources, as listed in Table \ref{tab:r0} and \ref{tab:omega}. 
Our results considered both errors of the statistic and the systematic, $R_{0}=7.92\pm0.16_{\rm{stat.}}\pm0.3_{\rm{sys.}}$~kpc and $\Omega_{\odot}=30.17\pm0.27_{\rm{stat.}}\pm0.3_{\rm{sys.}}$~km~s$^{-1}$~kpc$^{-1}$ are consistent with each other. 
Small difference in $R_{0}$ could be attributed to the employed Galactic rotation curve in the models and/or the different input samples as noted in the previous paragraph \citep{Honma2012, Honma2015}. 
As already discussed in previous papers, new VLBI astrometry results are different from those recommended by the International Astronomical Union (IAU), $R_{0}$=8.5~kpc and $\Theta_{0}$=220~km~s$^{-1}$ \citep{Kerr1986}. 
Our results yield $\Theta_{0}=R_{0}\Omega_{0}=227$~km~s$^{-1}$. 
Thus, $R_{0}$ and $\Theta_{0}$ are smaller by 6\% and larger by 3\%, respectively. 
The angular velocity of the Sun, $\Omega_{\odot}$ is independently determined by proper motion measurements of a supermassive black hole at the Galactic center, Sgr~A$^{*}$ \citep{Reid2004, Reid2020}.
As listed in Table \ref{tab:omega}, all the results of $\Omega_{\odot}$ are in good agreement. 
Recently, the distance to Sgr~A$^{*}$ is accurately determined to be $R_{0}=8.178\pm0.013_{\rm{stat.}}\pm0.022_{\rm{sys.}}$~kpc \citep{Gravity2019} and $R_{0}=7.946\pm0.050_{\rm{stat.}}\pm0.032_{\rm{sys.}}$~kpc \citep{Do2019} by measurements of stellar orbital motions around Sgr~A$^{*}$ (Table \ref{tab:r0}). 
Our result is consistent with them. 
This consistency suggests that Sgr~A$^{*}$ is truly located at the dynamical center of the Galactic rotation of LSR.
When we adopt $R_{0}=8.178$~kpc \citep{Gravity2019}, the Galactic rotation velocity is $\sim$3\% upwardly revised to 234~km~s$^{-1}$. 

\section{Summary and future prospects}

We have compiled all the astrometry measurements from VERA to construct the first version of the VERA catalogue. 
In total 99 target sources are listed in the catalogue including 21 new measurements in the present paper. 
The results are basically consistent with those from other VLBI astrometry project with VLBA (BeSSeL) and EVN while significant differences are also reported for several sources. 
It is mainly affected by the spatial structures of the target maser features and their time variation, along with the systematic calibration errors. 
The effects are more significant for the nearby lower-mass YSOs and AGBs. 

Using all the available VLBI astrometry data base, we model the Galactic structure to estimate the fundamental parameters such as the distance toward the Galactic center, $R_{0}$, angular velocity of the LSR around the Galactic center, $\Omega_{0}$, and the model of the rotation curve. 
Using these parameters, the angular velocity of the Sun, $\Omega_{\odot}$, is calculated to compare with the other results. 
The results, $R_{0}=7.92\pm0.16_{\rm{stat.}}\pm0.3_{\rm{sys.}}$~kpc and $\Omega_{\odot}=30.17\pm0.27_{\rm{stat.}}\pm0.3_{\rm{sys.}}$~km~s$^{-1}$~kpc$^{-1}$ are also consistent with those from VLBA \citep{Reid2019}. 

Further astrometry observations with VERA will be able to advance the studies on the Galaxy model by increasing the number of target sources along with reducing systematic errors due to insufficient samples \citep{Honma2015}. 
New VLBI array such as KaVA/EAVN and those in southern hemisphere, LBA \citep{Krishnan2015, Krishnan2017} and future SKA (Square Kilometer Array) in the VLBI mode \citep{Green2015} will improve astrometry accuracy for spatially extended sources and southern sources, respectively, which are still insufficient for the currently available VERA catalogue. 
The new data analysis tool, VEDA, will provide systematic astrometry results for future VERA observational data and reanalysis of previous archive data \citep{Nagayama2020b}.

\begin{ack}
We are grateful to the referee for valuable  comments to improve the manuscript. 
We would like to thank all the staff of Mizusawa VLBI Observatory of NAOJ to operate the VERA array, correlate the VLBI data, and analyze the results by using VEDA. 
We also thank staff and students of Kagoshima University VERA Group who involved in telescope operation. 
T. Hirota is financially supported by the MEXT/JSPS KAKENHI Grant Number 17K05398. 
HI is supported by the MEXT/JSPS KAKENHI (16H02167). 
J.O.C. acknowledges support by the Italian Ministry of Foreign Affairs and International Cooperation (MAECI Grant Number ZA18GR02) and the South African Department of Science and Technology’s National Research Foundation (DST-NRF Grant Number 113121) as part of the ISARP RADIOSKY2020 Joint Research Scheme. 
Data analysis were in part carried out on common use data analysis computer system at the Astronomy Data Center, ADC, of the National Astronomical Observatory of Japan. 
\end{ack}

\clearpage

\clearpage

\begin{landscape}
\setlength{\topmargin}{50mm}
\setlength{\textheight}{160mm}
\setlength{\tabcolsep}{3pt}
\begin{tiny}
\begin{longtable}{lllccllrrrrrrcl}
\caption{Astrometry results from VERA
\label{tab:vera}}
\hline
\hline
     & RA(J2000) & Dec(J2000) 
 &  $l$ & $b$ & $\pi_{\rm{VERA}}$ & $\sigma_{\rm{VERA}}$ 
  & $\mu_{x}$ & $\Delta \mu_{x}$ & $\mu_{y}$ & $\Delta \mu_{y}$ & $v_{lsr}$ & $\Delta v_{lsr}$ &      &  \\
NAME & (h:m:s) & (d:m:s) &  (deg)  & (deg) 
  & \multicolumn{2}{c}{(mas)} & \multicolumn{2}{c}{(mas~yr$^{-1}$)}  & \multicolumn{2}{c}{(mas~yr$^{-1}$)}
   & \multicolumn{2}{c}{(km~s$^{-1}$)} &  Type & Reference \\
\hline
\endhead
SY~Scl            & 00 07 36.2476 & $-$25 29 40.028 & 039.91 & $-$80.04 & 0.75  & 0.03  &  $+$5.57 & 0.04 &  $-$7.32 & 0.12 & $+$22.0 & 5.0 & AGB & \citet{Nyu2011} \\
IRAS~00259$+$5625 & 00 28 43.5075 & $+$56 41 56.868 & 119.80 & $-$06.03 & 0.412 & 0.123 &  $-$2.48 & 0.32 &  $-$2.85 & 0.65 & $-$38.3 & 3.1 & SFR & \citet{Sakai2014} \\
NGC~281           & 00 52 24.7008 & $+$56 33 50.527 & 123.06 & $-$06.30 & 0.355 & 0.030 &  $-$2.63 & 0.05 &  $-$1.86 & 0.08 & $-$30.0 & 5.0 & SFR & \citet{Sato2008} \\
G125.51$+$02.03   & 01 15 40.8027 & $+$64 46 40.766 & 125.51 & $+$02.03 & 0.145 & 0.023 &  $-$1.20 & 0.21 &  $-$0.33 & 0.27 & $-$57.0 & 9.0 & SFR & \citet{Koide2019}, \\
                  &               &                 &        &          &       &       &          &      &          &      &         &     &     &  \citet{Sakai2020b}$^{a}$ \\
W3(H$_{2}$O)      & 02 27 04.6800 & $+$61 52 24.566 & 133.94 & $+$01.06 & 0.527 & 0.016 &  $+$0.27 & 0.29 &  $-$1.24 & 0.15 & $-$55.6 & 1.2 & SFR & \citet{Matsumoto2011},  \\
                  &               &                 &        &          &       &       &          &      &          &      &         &     &     &  \citet{Nagayama2020b}$^{a}$ \\
G135.28$+$02.80   & 02 43 28.5825 & $+$62 57 08.390 & 135.28 & $+$02.80 & 0.124 & 0.011 &  $-$0.45 & 0.20 &  $+$0.09 & 0.16 & $-$72.9 & 1.6 & SFR & \citet{Nagayama2020a} \\
G137.07$+$03.00   & 02 58 13.1793 & $+$62 20 32.915 & 137.07 & $+$03.00 & 0.187 & 0.016 &  $-$0.57 & 0.16 &  $-$0.01 & 0.16 & $-$50.1 & 0.4 & SFR & \citet{Nagayama2020a} \\
L1448C            & 03 25 38.8784 & $+$30 44 05.252 & 157.57 & $-$21.94 & 4.31  & 0.33  & $+$21.90 & 0.70 & $-$23.10 & 3.30 &  $+$5.0 & 5.0 & SFR & \citet{Hirota2011} \\
NGC~1333~SVS13    & 03 29 03.7247 & $+$31 16 03.802 & 158.35 & $-$20.56 & 4.25  & 0.32  & $+$14.25 & 2.58 &  $-$9.95 & 0.74 &  $+$8.0 & 5.0 & SFR & \citet{Hirota2008a} \\
V637~Per          & 03 54 02.2577 & $+$36 32 17.926 & 159.10 & $-$13.20 & 0.94  & 0.02  &  $-$0.61 & 0.43 &  $-$0.90 & 0.37 & $-$97.8 & 0.9 & AGB & Present paper \\
L1482             & 04 30 27.4008 & $+$35 09 17.649 & 165.47 & $-$09.05 & 1.879 & 0.096 &  $+$3.07 & 0.06 &  $-$8.60 & 0.04 &  $+$1.0 & 5.0 & SFR & \citet{Omodaka2020} \\
BX~Eri            & 04 40 32.7762 & $-$14 12 02.710 & 211.48 & $-$35.33 & 2.116 & 0.105 &  $+$6.77 & 0.35 & $-$10.79 & 0.25 &  $-$0.3 & 0.1 & AGB & Present paper \\
T~Lep             & 05 04 50.8430 & $-$21 54 16.505 & 222.67 & $-$32.71 & 3.06  & 0.04  & $+$14.60 & 0.50 & $-$35.43 & 0.79 & $-$27.6 & 5.0 & AGB & \citet{Nakagawa2014} \\
IRAS~05137$+$3919 & 05 17 13.7410 & $+$39 22 19.880 & 168.06 & $+$00.82 & 0.086 & 0.027 &  $+$0.30 & 0.27 &  $-$0.89 & 0.73 & $-$27.0 & 5.0 & SFR & \citet{Honma2011} \\
BW~Cam            & 05 19 52.1643 & $+$63 15 54.684 & 143.43 & $+$20.09 & 0.749 & 0.189 &  $+$7.55 & 1.19 & $-$19.63 & 0.81 & $+$42.0 & 0.7 & AGB & Present paper \\
IRAS~05168$+$3634 & 05 20 22.0700 & $+$36 37 56.630 & 170.66 & $-$00.25 & 0.532 & 0.053 &  $+$0.23 & 1.07 &  $-$3.14 & 0.28 & $-$15.5 & 1.9 & SFR & \citet{Sakai2012} \\
AFGL~5142         & 05 30 48.0173 & $+$33 47 54.568 & 174.20 & $-$00.07 & 0.467 & 0.010 &  $+$0.32 & 0.27 &  $-$0.22 & 0.47 &  $-$2.0 & 5.0 & SFR & \citet{Burns2017} \\
Orion~KL          & 05 35 14.5050 & $-$05 22 30.450 & 209.00 & $-$19.38 & 2.39  & 0.03  &  $+$9.56 & 0.10 &  $-$3.83 & 0.15 &  $+$3.0 & 5.0 & SFR & \citet{Hirota2007},   \\
                  &               &                 &        &          &       &       &          &      &          &      &         &     &     & \citet{Kim2008}$^{a}$, \\
                  &               &                 &        &          &       &       &          &      &          &      &         &     &     & \citet{Nagayama2020b} \\
WB~673            & 05 38 00.3500 & $+$35 58 58.400 & 173.17 & $+$02.36 & 0.590 & 0.043 &  $+$0.01 & 0.03 &  $-$3.40 & 0.09 & $-$10.4 & 0.2 & SFR & Present paper \\
RW~Lep            & 05 38 52.7260 & $-$14 02 27.180 & 217.78 & $-$22.30 & 1.62  & 0.16  & $+$15.80 & 2.10 & $-$31.20 & 2.10 & $-$59.0 & 1.0 & AGB & \citet{Kamezaki2014b} \\
L1641~S3          & 05 39 56.0431 & $-$07 30 27.988 & 211.57 & $-$19.29 & 2.114 & 0.127 & $-$11.68 & 0.67 &  $-$7.74 & 0.36 &  $+$6.8 & 4.2 & SFR & Present paper \\
S235AB~MIR        & 05 40 53.3800 & $+$35 41 48.500 & 173.72 & $+$02.70 & 0.639 & 0.033 &  $+$0.08 & 0.12 &  $-$2.41 & 0.14 & $-$17.0 & 5.0 & SFR & \citet{Burns2015} \\
B35               & 05 44 29.2483 & $+$09 08 52.121 & 196.93 & $-$10.40 & 1.98  & 0.25  &  $-$2.30 & 0.53 &  $-$5.31 & 0.59 & $+$12.0 & 1.0 & SFR & Present paper \\
BX~Cam            & 05 46 44.3251 & $+$69 58 24.408 & 143.43 & $+$20.09 & 1.73  & 0.03  & $+$13.48 & 0.14 & $-$34.30 & 0.18 &   0.0 & 5.0 & AGB & \citet{Matsuno2020} \\
G192.16$-$03.81   & 05 58 13.5300 & $+$16 31 58.900 & 192.16 & $-$03.81 & 0.66  & 0.04  &  $+$0.69 & 0.15 &  $-$1.57 & 0.15 &  $+$5.7 & 5.0 & SFR & \citet{Shiozaki2011} \\
IRAS~06058$+$2138 & 06 08 53.4938 & $+$21 38 30.741 & 188.94 & $+$00.88 & 0.569 & 0.034 &  $+$1.06 & 0.18 &  $-$2.77 & 0.34 &  $+$3.0 & 5.0 & SFR & \citet{Oh2010} \\
IRAS~06061$+$2151 & 06 09 06.9746 & $+$21 50 41.405 & 188.79 & $+$01.03 & 0.496 & 0.031 &  $-$0.10 & 0.10 &  $-$3.91 & 0.07 &  $-$1.6 & 0.2 & SFR & \citet{Niinuma2011} \\
HH~12-15          & 06 10 50.1400 & $-$06 11 45.600 & 213.88 & $-$11.84 & 1.61  & 0.36  &  $-$0.36 & 1.68 &  $+$3.17 & 0.47 & $+$11.3 & 2.0 & SFR & Present paper \\
S255~IR-SMA1      & 06 12 54.0064 & $+$17 59 22.959 & 192.60 & $-$00.05 & 0.563 & 0.036 &  $-$0.13 & 0.20 &  $-$0.06 & 0.27 &  $+$5.3 & 5.0 & SFR & \citet{Burns2016} \\
S269              & 06 14 37.0800 & $+$13 49 36.700 & 196.45 & $-$01.67 & 0.189 & 0.008 &  $-$0.42 & 0.20 &  $-$0.12 & 0.20 & $+$18.0 & 5.0 & SFR & \citet{Honma2007}$^{a}$, \\
                  &               &                 &        &          &       &       &          &      &          &      &         &     &     &  \citet{Asaki2014} \\
G200.08$-$01.63   & 06 21 47.5742 & $+$10 39 22.811 & 200.08 & $-$01.63 & 0.200 & 0.017 &  $+$0.32 & 0.14 &  $-$0.14 & 0.16 & $+$36.3 & 0.6 & SFR & \citet{Nagayama2020a} \\
U~Lyn             & 06 40 46.4853 & $+$59 52 01.490 & 155.66 & $+$21.94 & 1.27  & 0.06  &  $+$0.80 & 0.57 &  $-$6.00 & 0.56 & $-$13.0 & 3.0 & AGB & \citet{Kamezaki2016a} \\
NGC~2264          & 06 41 09.8600 & $+$09 29 14.700 & 203.32 & $+$02.05 & 1.356 & 0.098 &  $-$1.08 & 0.58 &  $-$5.92 & 3.06 &  $+$7.0 & 3.0 & SFR & \citet{Kamezaki2014a} \\
WB~886            & 06 47 13.3000 & $+$00 26 05.920 & 212.06 & $-$00.74 & 0.349 & 0.052 &  $-$0.40 & 0.94 &  $+$0.37 & 0.33 & $+$45.0 & 3.0 & SFR & \citet{Nakanishi2020} \\
NSV~17351         & 07 07 49.3869 & $-$10 44 05.998 & 224.34 & $-$01.29 & 0.247 & 0.010 &  $-$1.19 & 0.11 &  $+$1.30 & 0.19 & $-$50.1 & 1.9 & AGB & \citet{Morita2020} \\
VY~CMa            & 07 22 58.3291 & $-$25 46 03.141 & 239.35 & $-$05.06 & 0.88  & 0.08  &  $-$2.09 & 0.16 &  $+$1.02 & 0.61 & $+$20.0 & 5.0 & RSG & \citet{Choi2008} \\
OZ~Gem            & 07 33 57.7500 & $+$30 30 37.799 & 188.80 & $+$21.90 & 0.806 & 0.039 &  $-$1.97 & 0.32 &  $-$8.69 & 0.21 &  $+$8.7 & 1.4 & AGB & \citet{Urago2020} \\
QX~Pup            & 07 42 16.9470 & $-$14 42 50.200 & 231.84 & $+$04.22 & 0.61  & 0.03  &  $-$4.76 & 0.37 &  $-$0.94 & 0.62 & $+$33.0 & 5.0 & AGB & \citet{Ooyama2020} \\
IRAS~07427$-$2400 & 07 44 51.9200 & $-$24 07 41.500 & 240.31 & $+$00.07 & 0.185 & 0.027 &  $-$1.79 & 0.32 &  $+$2.60 & 0.17 & $+$66.4 & 5.0 & SFR & \citet{Sakai2015} \\
HU~Pup            & 07 55 40.1843 & $-$28 38 54.608 & 245.44 & $-$00.15 & 0.308 & 0.042 &  $-$1.16 & 0.15 &  $+$3.69 & 0.20 & $+$43.9 & 0.6 & AGB & Present paper \\
R~Cnc             & 08 16 33.8243 & $+$11 43 34.518 & 211.75 & $+$24.14 & 3.84  & 0.29  &  $+$1.24 & 0.34 & $-$11.57 & 0.97 & $+$16.9 & 5.0 & AGB & Present paper \\
X~Hya             & 09 35 30.2650 & $-$14 41 28.639 & 248.15 & $+$26.70 & 2.07  & 0.05  & $-$51.37 & 0.97 & $-$15.02 & 1.47 & $+$27.3 & 5.0 & AGB & Present paper \\
R~UMa             & 10 44 38.4283 & $+$68 46 32.344 & 138.36 & $+$44.36 & 1.97  & 0.05  & $-$40.77 & 0.39 & $-$24.75 & 0.38 & $+$40.5 & 1.0 & AGB & \citet{Nakagawa2016} \\
W~Leo             & 10 53 37.4325 & $+$13 42 54.367 & 233.02 & $+$59.43 & 1.03  & 0.02  &  $-$6.84 & 0.09 &  $-$8.65 & 0.08 & $+$46.7 & 0.2 & AGB & Present paper \\
HS~UMa            & 11 35 30.6878 & $+$34 52 04.006 & 182.78 & $+$72.02 & 2.816 & 0.095 & $-$11.48 & 0.17 & $-$10.86 & 0.65 &  $+$1.6 & 0.3 & AGB & Present paper \\
S~Crt             & 11 52 45.9697 & $-$07 35 48.096 & 278.59 & $+$52.48 & 2.33  & 0.13  &  $-$3.17 & 0.22 &  $-$5.41 & 0.22 & $+$37.9 & 5.0 & AGB & \citet{Nakagawa2008} \\
R~Hya             & 13 29 42.7819 & $-$23 16 52.775 & 314.22 & $+$38.75 & 7.93  & 0.18  & $-$53.79 & 1.05 & $+$16.15 & 1.83 &  $-$8.5 & 5.0 & AGB & Present paper \\
RX~Boo            & 14 24 11.6206 & $+$25 42 12.909 & 034.28 & $+$69.21 & 7.31  & 0.50  & $+$24.55 & 1.06 & $-$49.67 & 2.38 &  $+$1.0 & 5.0 & AGB & \citet{Kamezaki2012} \\
FV~Boo            & 15 08 25.7530 & $+$09 36 18.390 & 011.03 & $+$53.27 & 0.97  & 0.06  &  $+$6.81 & 0.14 &  $+$1.01 & 0.12 &  $+$7.5 & 1.0 & AGB & \citet{Kamezaki2016b} \\
Y~Lib             & 15 11 41.2990 & $-$06 00 41.462 & 353.83 & $+$42.59 & 0.855 & 0.050 & $-$10.15 & 2.39 & $-$15.02 & 4.26 & $+$14.4 & 1.1 & AGB & \citet{Chibueze2019} \\
S~Ser             & 15 21 39.5334 & $+$14 18 53.107 & 020.50 & $+$52.79 & 1.25  & 0.04  &  $-$2.56 & 1.42 &  $+$5.20 & 2.31 & $+$25.1 & 5.0 & AGB & Present paper \\
IRAS~16293$-$2422 & 16 32 22.8500 & $-$24 28 36.400 & 353.94 & $-$15.84 & 5.6   & 1.5   & $-$20.60 & 0.70 & $-$32.40 & 2.00 &  $+$3.0 & 5.0 & SFR & \citet{Imai2007} \\
NGC~6334I(N)      & 17 20 55.1920 & $-$35 45 03.770 & 351.44 & $+$00.65 & 0.789 & 0.161 &  $-$2.88 & 0.30 &  $+$3.23 & 0.39 &  $-$2.8 & 0.5 & SFR & \citet{Chibueze2014a} \\
G353.27$+$00.64   & 17 26 01.5883 & $-$34 15 14.903 & 353.27 & $+$00.64 & 0.59  & 0.06  &  $+$0.47 & 0.07 &  $+$0.99 & 1.04 &  $-$5.0 & 5.0 & SFR & \citet{Motogi2016} \\
G359.62$-$00.25   & 17 45 39.0908 & $-$29 20 26.294 & 359.62 & $-$00.25 & 0.33  & 0.14  &  $+$1.31 & 0.33 &  $-$2.41 & 0.87 & $-$80.0 & 5.0 & SFR & \citet{Iwata2017} \\
Sgr~B2            & 17 47 20.1817 & $-$28 23 03.889 & 000.67 & $-$00.03 & 0.133 & 0.038 &  $-$1.83 & 0.21 &  $-$3.70 & 0.09 & $+$62.0 & 5.0 & SFR & \citet{Sakai2020a} \\
Sgr~D             & 17 48 48.5450 & $-$28 01 26.290 & 001.15 & $-$00.12 & 0.423 & 0.083 &  $-$0.76 & 0.15 &  $-$2.88 & 0.34 & $-$18.0 & 5.0 & SFR & \citet{Sakai2017} \\
G005.88$-$00.39   & 18 00 30.3100 & $-$24 04 04.500 & 005.88 & $-$00.39 & 0.78  & 0.05  &  $-$0.17 & 0.60 &  $-$0.95 & 0.48 &  $+$9.0 & 3.0 & SFR & \citet{Motogi2011} \\
G007.47$+$00.06   & 18 02 13.1790 & $-$22 27 58.960 & 007.47 & $+$00.06 & --- & --- &  $-$2.42 & 0.09 &  $-$4.39 & 0.08 & $-$15.0 & 5.0 & SFR & \citet{Yamauchi2016}$^{b}$ \\
G014.33$-$00.64   & 18 18 54.6532 & $-$16 47 50.077 & 014.33 & $-$00.64 & 0.893 & 0.101 &  $+$0.95 & 2.00 &  $-$2.50 & 2.00 & $+$22.0 & 5.0 & SFR & \citet{Sato2010} \\
M17               & 18 20 23.0160 & $-$16 11 48.030 & 015.03 & $-$00.67 & 0.491 & 0.041 &  $-$0.51 & 0.21 &  $-$2.04 & 0.21 & $+$20.0 & 5.0 & SFR & \citet{Chibueze2016} \\
G021.88$+$00.02   & 18 31 01.7490 & $-$09 49 01.130 & 021.88 & $+$00.01 & --- & --- &  $-$3.30 & 0.06 &  $-$5.33 & 0.22 & $+$26.9 & 0.4 & SFR & Present paper$^{b}$ \\
IRAS~18286$-$0959 & 18 31 22.9340 & $-$09 57 21.700 & 021.80 & $-$00.13 & 0.277 & 0.041 &  $-$3.20 & 0.30 &  $-$7.20 & 0.20 & $+$60.0 & 5.0 & AGB & \citet{Imai2013} \\
G034.39$+$00.22   & 18 53 18.7700 & $+$01 24 08.800 & 034.39 & $+$00.22 & 0.643 & 0.049 &  $-$0.25 & 0.80 &  --- & --- & $+$58.0 & 5.0 & SFR & \citet{Kurayama2011}$^{b}$ \\
S76E              & 18 56 11.4413 & $+$07 53 17.608 & 040.50 & $+$02.54 & 0.521 & 0.024 &  $-$0.89 & 0.34 &  $-$2.27 & 0.56 & $+$31.9 & 1.7 & SFR & \citet{Chibueze2017} \\
G037.50$+$00.53   & 18 57 53.3876 & $+$04 18 17.394 & 037.50 & $+$00.53 & 0.091 & 0.016 &  $-$2.74 & 0.18 &  $-$5.49 & 0.10 & $+$10.7 & 2.6 & SFR & \citet{Nagayama2020a} \\
G037.82$+$00.41   & 18 58 53.8800 & $+$04 32 15.004 & 037.82 & $+$00.41 & 0.089 & 0.020 &  $-$2.73 & 0.12 &  $-$5.53 & 0.12 & $+$17.5 & 0.8 & SFR & \citet{Nagayama2020a} \\
W48A              & 19 01 45.5423 & $+$01 13 32.573 & 035.20 & $-$01.74 & 0.433 & 0.026 &  $-$0.05 & 0.81 &  $-$3.51 & 0.38 & $+$41.9 & 1.4 & SFR & \citet{Chibueze2020} \\
G044.31$+$00.04   & 19 12 15.7930 & $+$10 07 53.085 & 044.31 & $+$00.04 & 0.192 & 0.031 &  $-$3.36 & 0.05 &  $-$6.92 & 0.06 & $+$57.8 & 0.5 & SFR & Present paper \\
G048.60$+$00.02   & 19 20 31.1772 & $+$13 55 25.257 & 048.60 & $+$00.02 & 0.199 & 0.007 &  $-$2.76 & 0.04 &  $-$5.28 & 0.11 & $+$19.0 & 1.0 & SFR & \citet{Nagayama2011a} \\
G048.99$-$00.30   & 19 22 26.1348 & $+$14 06 39.133 & 048.99 & $-$00.30 & 0.178 & 0.017 &  $-$2.16 & 0.09 &  $-$5.87 & 0.17 & $+$66.3 & 0.3 & SFR & \citet{Nagayama2015a} \\
G049.19$-$00.33   & 19 22 57.7705 & $+$14 16 09.969 & 049.19 & $-$00.33 & 0.211 & 0.016 &  $-$3.21 & 0.07 &  $-$5.08 & 0.25 & $+$69.9 & 0.5 & SFR & \citet{Nagayama2015a} \\
IRAS~19213$+$1723 & 19 23 37.3229 & $+$17 29 10.479 & 052.10 & $+$01.04 & 0.251 & 0.036 &  $-$2.53 & 0.04 &  $-$6.07 & 0.05 & $+$41.7 & 5.0 & SFR & \citet{Oh2010} \\
K3-35             & 19 27 44.0230 & $+$21 30 03.440 & 056.10 & $+$02.09 & 0.260 & 0.040 &  $-$3.34 & 0.10 &  $-$5.93 & 0.07 & $+$26.0 & 5.0 & AGB & \citet{Tafoya2011} \\
IRAS~19312$+$1950 & 19 33 24.2430 & $+$19 56 55.650 & 055.37 & $+$00.19 & 0.263 & 0.047 &  $-$2.61 & 0.47 &  $-$6.73 & 0.14 & $+$36.0 & 1.0 & SFR & \citet{Imai2011} \\
G061.48$+$00.10   & 19 46 47.9175 & $+$25 12 52.698 & 061.48 & $+$00.10 & 0.454 & 0.022 &  $-$1.31 & 0.16 &  $-$6.39 & 0.34 & $+$41.7 & 6.2 & SFR & Present paper \\
SY~Aql            & 20 07 05.4083 & $+$12 57 06.219 & 053.37 & $-$10.31 & 1.10  & 0.07  & $+$12.26 & 0.11 & $-$15.93 & 0.22 & $-$44.8 & 5.0 & AGB & Present paper \\
IRAS~20056$+$3350 & 20 07 31.2586 & $+$33 59 41.477 & 071.31 & $+$00.83 & 0.213 & 0.026 &  $-$2.62 & 0.33 &  $-$5.65 & 0.52 &  $+$9.4 & 5.0 & SFR & \citet{Burns2014a} \\
ON1               & 20 10 09.2045 & $+$31 31 36.101 & 069.54 & $-$00.97 & 0.404 & 0.017 &  $-$3.10 & 0.18 &  $-$4.70 & 0.24 & $+$12.0 & 1.0 & SFR & \citet{Nagayama2011b} \\
IRAS~20126$+$4104 & 20 14 26.0218 & $+$41 13 32.674 & 078.12 & $+$03.63 & 0.750 & 0.092 &  $-$4.15 & 0.51 &  $-$4.07 & 0.51 &  $-$3.5 & 4.0 & SFR & \citet{Nagayama2015b} \\
IRAS~20143$+$3634 & 20 16 13.3617 & $+$36 43 33.920 & 074.57 & $+$00.85 & 0.367 & 0.037 &  $-$2.99 & 0.16 &  $-$4.37 & 0.43 &  $-$1.0 & 1.0 & SFR & \citet{Burns2014b} \\
ON2N              & 20 21 44.0123 & $+$37 26 37.484 & 075.78 & $+$00.34 & 0.261 & 0.009 &  $-$2.79 & 0.13 &  $-$4.66 & 0.17 &  $+$0.0 & 1.0 & SFR & \citet{Ando2011} \\
IRAS~20231$+$3430 & 20 25 07.8013 & $+$34 50 34.733 & 074.04 & $-$01.71 & 0.611 & 0.022 &  $-$3.79 & 0.18 &  $-$4.88 & 0.25 &  $+$6.0 & 5.0 & SFR & \citet{Ogbodo2017} \\
IRAS~20255$+$4032 & 20 27 20.2734 & $+$40 42 34.648 & 079.09 & $+$01.33 & 0.118 & 0.035 &  $-$2.49 & 0.13 &  $-$3.36 & 0.23 & $-$18.2 & 5.0 & SFR & \citet{Sakai2020c} \\
G080.70$+$00.70   & 20 35 09.1650 & $+$41 38 20.260 & 080.70 & $+$00.70 & 0.258 & 0.022 &  $-$3.18 & 0.09 &  $-$5.09 & 0.07 &  $-$2.3 & 0.7 & SFR & Present paper \\
G095.05$+$03.97   & 21 15 55.6798 & $+$54 43 31.328 & 095.05 & $+$03.97 & 0.108 & 0.023 &  $-$2.44 & 0.21 &  $-$2.63 & 0.17 & $-$87.0 & 5.0 & SFR & \citet{Sakai2020b}$^{a}$, \\
                  &               &                 &        &          &       &       &          &      &          &      &         &     &     & \citet{Nakanishi2020} \\
G097.53$+$03.18   & 21 32 12.4400 & $+$55 53 49.600 & 097.53 & $+$03.18 & 0.177 & 0.028 &  $-$2.64 & 0.20 &  $-$2.38 & 0.22 & $-$73.0 & 5.0 & SFR & \citet{Sakai2020b}$^{a}$, \\
                  &               &                 &        &          &       &       &          &      &          &      &         &     &     & \citet{Nakanishi2020} \\
IRAS~21379$+$5106 & 21 39 40.5500 & $+$51 20 34.000 & 095.29 & $-$00.93 & 0.262 & 0.031 &  $-$2.74 & 0.08 &  $-$2.87 & 0.18 & $-$42.3 & 0.2 & SFR & \citet{Nakanishi2015} \\
AFGL~2789         & 21 39 58.2717 & $+$50 14 21.014 & 094.60 & $-$01.79 & 0.326 & 0.031 &  $-$2.20 & 0.08 &  $-$3.77 & 0.15 & $-$44.0 & 5.0 & SFR & \citet{Oh2010} \\
G102.35$+$03.64   & 21 57 25.1841 & $+$59 21 56.614 & 102.35 & $+$03.64 & 0.154 & 0.021 &  $-$2.53 & 0.33 &  $-$2.14 & 0.33 & $-$88.0 & 5.0 & SFR & \citet{Sakai2020b}$^{a}$, \\
                  &               &                 &        &          &       &       &          &      &          &      &         &     &     & \citet{Nakanishi2020} \\
SV~Peg            & 22 05 42.0850 & $+$35 20 54.536 & 088.72 & $-$16.29 & 3.00  & 0.06  & $+$11.59 & 0.54 &  $-$8.63 & 0.44 &  $+$3.9 & 5.0 & AGB & \citet{Sudou2019} \\
S140              & 22 19 17.4657 & $+$63 18 39.851 & 106.79 & $+$05.31 & 1.154 & 0.069 &  $-$6.16 & 0.12 &  $-$4.74 & 0.11 &  $-$6.1 & 5.0 & SFR & Present paper \\
IRAS~22198$+$6336 & 22 21 26.7279 & $+$63 51 37.924 & 107.29 & $+$05.63 & 1.309 & 0.047 &  $-$2.47 & 1.40 &  $+$0.26 & 1.40 & $-$11.0 & 5.0 & SFR & \citet{Hirota2008b} \\
IRAS~22480$+$6002 & 22 49 58.8760 & $+$60 17 56.650 & 108.43 & $-$00.89 & 0.400 & 0.025 &  $-$2.58 & 0.33 &  $-$1.91 & 0.17 & $-$50.8 & 3.5 & SFR & \citet{Imai2012} \\
IRAS~22555$+$6213 & 22 57 29.8090 & $+$62 29 46.850 & 110.20 & $+$02.48 & 0.314 & 0.070 &  $-$2.04 & 0.05 &  $-$0.66 & 0.06 & $-$63.0 & 1.0 & SFR & \citet{Chibueze2014b} \\
IRAS~23004$+$5642 & 23 02 32.0800 & $+$56 57 51.400 & 108.47 & $-$02.81 & 0.309 & 0.010 &  $-$2.45 & 1.00 &  $-$3.00 & 0.70 & $-$54.0 & 5.0 & SFR & \citet{Nakanishi2020} \\
R~Peg             & 23 06 39.1652 & $+$10 32 36.078 & 085.41 & $-$44.56 & 2.76  & 0.28  &  $+$3.60 & 1.53 &  $-$6.44 & 0.92 & $+$22.5 & 5.0 & AGB & Present paper \\
R~Aqr             & 23 43 49.4616 & $-$15 17 04.202 & 066.52 & $-$70.33 & 4.59  & 0.24  & $+$37.13 & 0.47 & $-$28.62 & 0.44 & $-$21.5 & 5.0 & AGB & \citet{Kamohara2010}, \\
                  &               &                 &        &          &       &       &          &      &          &      &         &     &     & \citet{Min2014}$^{a}$\\
PZ~Cas            & 23 44 03.2816 & $+$61 47 22.187 & 115.06 & $-$00.05 & 0.356 & 0.026 &  $-$3.70 & 0.20 &  $-$2.00 & 0.30 & $-$36.2 & 0.7 & RSG & \citet{Kusuno2013} \\
\hline
\multicolumn{15}{l}{$a$: If there are multiple references, the data with smaller parallax errors noted with $^{a}$ is employed. } \\
\multicolumn{15}{l}{$b$: Their parallax and/or proper motions cannot be determined. } 
\end{longtable}
\end{tiny}
\end{landscape}
\renewcommand{\tabcolsep}{6pt} 

\begin{table*}[htb]
\tbl{Parallaxes from VERA and VLBA/EVN and their differences}{
\begin{tabular}{lcccrl} 
\hline
\hline
         & $\pi_{\rm{VERA}}$ & $\pi_{\rm{VLBA}}$ & $\pi_{\rm{VERA}}-\pi_{\rm{VLBA}}$  &  &  References for  \\
Name     &  (mas) &  (mas) &  (mas) & $\Delta \pi/\sigma_{\Delta \pi}$ &   VLBA/EVN \\
\hline
NGC~281/NGC~281-W    & 0.355$\pm$0.030  & 0.421$\pm$0.022  &   $-0.07\pm0.04$ &   $-1.77$ & \citet{Rygl2010} \\ 
W3(H$_{2}$O)/W3(OH)  & 0.527$\pm$0.016  & 0.512$\pm$0.010  &    $0.02\pm0.02$ &    $0.79$ & \citet{Xu2006}$^{a}$, \citet{Hachisuka2006} \\ 
G135.28$+$02.80      & 0.124$\pm$0.011  & 0.167$\pm$0.006  &   $-0.04\pm0.01$ &   $-3.43$ & \citet{Hachisuka2009} \\ 
IRAS~05137$+$3919    & 0.086$\pm$0.027  & 0.201$\pm$0.024  &   $-0.12\pm0.04$ &   $-3.18$ & \citet{Hachisuka2015} \\ 
Orion~KL             &  2.39$\pm$0.03   & 2.415$\pm$0.040  &   $-0.02\pm0.05$ &   $-0.50$ & \citet{Menten2007} \\ 
IRAS~06058$+$2138    & 0.569$\pm$0.034  & 0.476$\pm$0.006  &    $0.09\pm0.03$ &    $2.69$ & \citet{Reid2009a}$^{a}$, \citet{Sakai2019} \\
HH~12-15             &  1.61$\pm$0.36  &   1.12$\pm$0.05   &    $0.48\pm0.36$ &    $1.35$ & \citet{Dzib2016} \\ 
S255~IR-SMA1         & 0.563$\pm$0.036  & 0.628$\pm$0.027  &   $-0.07\pm0.04$ &   $-1.44$ & \citet{Rygl2010} \\ 
S269                 & 0.189$\pm$0.008  & 0.241$\pm$0.012  &   $-0.05\pm0.01$ &   $-3.61$ & \citet{Quiroga-Nunez2019} \\ 
VY~CMa               &  0.88$\pm$0.08   &  0.83$\pm$0.08   &    $0.05\pm0.11$ &    $0.44$ & \citet{Zhang2012} \\ 
IRAS~07427$-$2400    & 0.185$\pm$0.027  & 0.188$\pm$0.016  &   $-0.00\pm0.03$ &   $-0.10$ & \citet{Choi2014} \\ 
IRAS~16293$-$2422    &   5.6$\pm$1.5    &   7.1$\pm$1.3    &   $-1.50\pm1.98$ &   $-0.76$ & \citet{Dzib2018a} \\ 
NGC~6334I(N)         & 0.789$\pm$0.161  & 0.744$\pm$0.076  &    $0.05\pm0.18$ &    $0.25$ & \citet{Wu2014} \\ 
G359.62$-$00.25      &  0.33$\pm$0.14   & 0.375$\pm$0.021  &   $-0.04\pm0.14$ &   $-0.32$ & \citet{Reid2019} \\ 
Sgr~B2               & 0.133$\pm$0.038  & 0.129$\pm$0.012  &    $0.00\pm0.04$ &    $0.10$ & \citet{Reid2009c} \\ 
Sgr~D                & 0.423$\pm$0.083  & 0.194$\pm$0.161  &    $0.23\pm0.18$ &    $1.26$ & \citet{Reid2019} \\ 
G005.88$-$00.39      &  0.78$\pm$0.05   & 0.334$\pm$0.020  &    $0.45\pm0.05$ &    $8.28$ & \citet{Sato2014} \\ 
M17                  & 0.491$\pm$0.041  & 0.505$\pm$0.033  &   $-0.01\pm0.05$ &   $-0.27$ & \citet{Xu2011} \\ 
W48A                 & 0.433$\pm$0.026  & 0.306$\pm$0.045  &    $0.13\pm0.05$ &    $2.44$ & \citet{Zhang2009} \\
G048.60$+$00.02      & 0.199$\pm$0.007  & 0.093$\pm$0.005  &    $0.11\pm0.01$ &   $12.32$ & \citet{Zhang2013} \\ 
G049.19$-$00.33      & 0.211$\pm$0.016  & 0.192$\pm$0.009  &    $0.02\pm0.02$ &    $1.03$ & \citet{Wu2014} \\ 
IRAS~19213$+$1723    & 0.251$\pm$0.036  & 0.162$\pm$0.013  &    $0.09\pm0.04$ &    $2.33$ & \citet{Wu2019} \\ 
ON1                  & 0.404$\pm$0.017  & 0.425$\pm$0.036  &   $-0.02\pm0.04$ &   $-0.53$ & \citet{Rygl2010}, \citet{Xu2013}$^{a}$ \\
IRAS~20126$+$4104    & 0.750$\pm$0.092  &  0.61$\pm$0.02   &    $0.14\pm0.09$ &    $1.49$ & \citet{Moscadelli2011} \\ 
IRAS~20231$+$3430    & 0.611$\pm$0.022  & 0.629$\pm$0.017  &   $-0.02\pm0.03$ &   $-0.65$ & \citet{Xu2013} \\ 
G097.53$+$03.18      & 0.177$\pm$0.028  & 0.133$\pm$0.017  &    $0.04\pm0.03$ &    $1.34$ & \citet{Hachisuka2015} \\ 
IRAS~21379$+$5106    & 0.262$\pm$0.031  & 0.206$\pm$0.007  &    $0.06\pm0.03$ &    $1.76$ & \citet{Choi2014} \\ 
AFGL~2789            & 0.326$\pm$0.031  & 0.253$\pm$0.024  &    $0.07\pm0.04$ &    $1.86$ & \citet{Choi2014}$^{a}$, \citet{Sakai2019} \\
\hline
\multicolumn{6}{l}{${a}$ If there are multiple references, we refer to papers labeled with ${a}$ reporting higher accuracy parallaxes. }\\
\end{tabular}}
\label{tab:compare}
\end{table*}

\clearpage

\begin{table*}[htb]
\tbl{Estimated Galactic parameters}{
\begin{tabular}{lccccc}
\hline
\hline
                                      & \multicolumn{2}{c}{Power-law model}         & & \multicolumn{2}{c}{2nd-order polynomial model} \\
\cline{2-3} \cline{5-6} 
Parameter                             & Present study      & \citet{Honma2012} ID14 & & Present study  & \citet{Honma2012} ID22 \\
\hline
$R_0$ (kpc)                           & $7.92\pm0.16$      & $7.82\pm0.41$          & & $7.97\pm0.15$  & $7.70\pm0.40$          \\
$\Omega_0$ (km~s$^{-1}$~kpc$^{-1}$)   & $28.63\pm0.26$     & $29.60\pm0.74$         & & $28.64\pm0.26$ & $29.71\pm0.71$         \\
$U_{\rm{s}}$ (km~s$^{-1}$)            & $4.2\pm1.0$        & $0.8\pm1.4$            & & $4.3\pm1.0$    & $0.7\pm1.4$            \\
$V_{\rm{s}}$ (km~s$^{-1}$)            & $-4.9\pm0.9$       & $-6.3\pm1.2$           & & $-3.7\pm1.0$   & $-6.5\pm1.3$           \\
$W_{\rm{s}}$ (km~s$^{-1}$)            & $-0.4\pm0.7$       & $-1.9\pm1.1$           & & $-0.4\pm0.7$   & $-1.9\pm1.1$           \\
$\alpha$                              & $-0.016\pm0.012$   & 0.00$\pm$0.02          & & ---            & ---                    \\
$a_0$  (km s$^{-1}$ kpc$^{-1}$)       & ---                & ---                    & & $-0.5\pm0.4$   & $-0.1\pm0.7$           \\
$b_0$  (km s$^{-1}$ kpc$^{-2}$)       & ---                & ---                    & & $-0.2\pm0.1$   & $0.1\pm0.2$            \\
\hline
\end{tabular}}
\label{tab:prm}
\end{table*}

\begin{table*}[htb]
\tbl{Comparison of Galactic center distance $R_{0}$}{
\begin{tabular}{llc} 
\hline
\hline
Method                                     & Reference            & $R_{0}$ (kpc)             \\
\hline
VLBI astrometry of 188 maser sources       & Present work         & $7.92\pm0.16_{\rm{stat.}}\pm0.3_{\rm{sys.}}$     \\
VLBI astrometry of 147 maser sources       & \citet{Reid2019}     & 8.15$\pm$0.15             \\   
Orbital motion of S2 around Sgr~A$^{*}$    & \citet{Gravity2019}  & $8.178\pm0.013_{\rm{stat.}}\pm0.022_{\rm{sys.}}$ \\   
Orbital motions of S0-2 around Sgr~A$^{*}$ & \citet{Do2019}       & $7.946\pm0.050_{\rm{stat.}}\pm0.032_{\rm{sys.}}$ \\   
\hline
\end{tabular}}
\label{tab:r0}
\end{table*}

\begin{table*}[htb]
\tbl{Comparison of angular velocity of the Sun $\Omega_{\odot}$}{
\begin{tabular}{llc} 
\hline
\hline
Method                                     & Reference            & $\Omega_{\odot}$ (km~s$^{-1}$~kpc$^{-1}$) \\
\hline
VLBI astrometry of 188 maser sources       & Present work         & $30.17\pm0.27_{\rm{stat.}}\pm0.3_{\rm{sys.}}$    \\
VLBI astrometry of 147 maser sources       & \citet{Reid2019}     & 30.32$\pm$0.27            \\   
Proper motion of Sgr~A$^{*}$        & \citet{Reid2020}     & 30.39$\pm$0.04   	      \\
\hline
\end{tabular}}
\label{tab:omega}
\end{table*}

\end{document}